\documentclass[fleqn,usenatbib]{mnras}

\usepackage{newtxtext,newtxmath}

\usepackage[T1]{fontenc}

\DeclareRobustCommand{\VAN}[3]{#2}
\let\VANthebibliography\thebibliography
\def\thebibliography{\DeclareRobustCommand{\VAN}[3]{##3}\VANthebibliography}

\def\kms{km~s$^{-1}$}

\usepackage{graphicx}	
\usepackage{amsmath}	
\usepackage{multirow}
\usepackage[normalem]{ulem}

\title[glSNe in discoverable galaxy-galaxy lenses]{Find the haystacks, then look for needles: The rate of strongly lensed transients in galaxy-galaxy strong gravitational lenses}
\author[A. Sainz de Murieta et al.]{
Ana Sainz de Murieta$^{1}$\thanks{E-mail: ana.sainz-de-murieta@port.ac.uk},
Thomas E. Collett$^{1}$,
Mark R. Magee$^{2}$, 
Justin D. R. Pierel$^{3}$,
Wolfgang J. R. Enzi$^{1}$,
\newauthor Martine Lokken$^{4}$, Alex Gagliano$^{5,6}$ and Dan Ryczanowski$^{1}$,
\\ 
$^{1}$ Institute of Cosmology and Gravitation, University of Portsmouth, Burnaby Rd, Portsmouth PO1 3FX, UK\\
$^{2}$ Department of Physics, University of Warwick, Gibbet Hill Road, Coventry CV4 7AL, UK\\
$^{3}$ Space Telescope Science Institute, Baltimore, MD 21218, USA\\
$^{4}$ Institut de F\'{i}sica d'Altes Energies (IFAE), The Barcelona Institute of Science and Technology, Campus UAB, 08193 Bellaterra (Barcelona) Spain\\
$^{5}$ The NSF AI Institute for Artificial Intelligence and Fundamental Interactions\\
$^{6}$ Center for Astrophysics | Harvard \& Smithsonian, 60 Garden Street, Cambridge, MA 02138-1516, USA
}

\date{Accepted XXX. Received YYY; in original form ZZZ}

\pubyear{2024}

\begin{document}
\label{firstpage}
\pagerange{\pageref{firstpage}--\pageref{lastpage}}
\maketitle

\begin{abstract}
The time delay between appearances of multiple images of a gravitationally lensed supernova (glSN) is sensitive to the Hubble constant, $H_0$. As well as time delays, a lensed host galaxy is needed to enable precise inference of $H_0$. In this work we investigate the connection between discoverable lensed transients and their host galaxies. We find that LSST will discover 88 glSNe per year, of which 54\% will also have a strongly lensed host. The rates can change by approximately 30 percent uncertainty depending primarily on the choice of unlensed SN population and uncertainties in the redshift evolution of the deflector population, but the fraction of glSNe with a lensed host is consistently around a half. LSST will discover 20 glSNe per year in systems that could plausibly have been identified by Euclid as galaxy-galaxy lenses before the discovery of the glSN. Such systems have preferentially longer time delays and therefore are well suited for cosmography. We define a golden sample of glSNe~Ia with time delays over 10 days, image separations greater than 0.8 arcseconds, and a multiply imaged host. For this golden sample, we find $91\%$ occur in systems that should already be discoverable as galaxy-galaxy lenses in Euclid. For cosmology with glSNe, monitoring Euclid lenses is a plausible alternative to searching the entire LSST alert stream. 
\end{abstract}




\section{Introduction}
Strong gravitational lensing is the phenomenon by which the intense gravitational potential from a foreground galaxy or cluster deflects the light coming from a background source and causes multiple images to appear. It serves as a powerful probe in astrophysics and cosmology. This rare effect can provide valuable insights into high-redshift sources, the distribution of matter in large galaxies and clusters, and the expansion rate of the Universe \citep{2010ARA&A..48...87T}. Using strong gravitational lensing to measure the expansion rate of the Universe was first suggested by \cite{refsdal--64}, who proposed using the time difference between the appearance of lensed images of a supernova, combined with an accurate model of the lens potential, to measure the time delay distance and hence the expansion rate, or the Hubble constant ($H_0$). Local measurements of $H_0$ combining Cepheids and Type Ia supernovae (SNe~Ia) yield a value of $73.30 \pm 1.04$ km~s$^{-1}$~Mpc$^{-1}$ \citep{Riess_2022}, and are in disagreement with "early-time" measurements from the Cosmic Microwave Background, for which the latest analysis yields a value of $H_0=67.4\pm 0.5$ km~s$^{-1}$~Mpc$^{-1}$ \citep{planck18--20}. Additional independent measurements, such as those from strong lensing time-delays, must be obtained if we are to test whether this tension is due to systematics or physics beyond our current cosmological model \citep{2021CQGra..38o3001D}.  

Gravitationally lensed quasars have dominated the field of time-delay cosmography ever since their discovery \citep{Walsh1979}. The discovery of numerous multiply-imaged quasars along with continuous light curve monitoring allowed for the first time delay measurements (e.g. \citealt{Kundić_1997}; \citealt{Schechter_1997}). Subsequent advancements in both light curve analysis and lens modelling throughout the 21st century have yielded a value of $H_0=73.3^{+1.7}_{-1.8}$ km s$^{-1}$ Mpc$^{-1}$ (\citealt{wong--20}, \citealt{millon--20}) from 6 lensed quasars, consistent with the local Cepheid and SNe~Ia measurement. 

Despite their maturity as a cosmological probe, gravitationally lensed quasars pose certain challenges. Firstly, their light curves are stochastic and require years of monitoring to measure precise time delays. Secondly, they are consistently bright, outshining their host and making it hard to use the morphology of the host as constraints on the lens model. These challenges can be overcome by using gravitationally lensed supernovae (glSNe) as they vary on shorter timescales and they fade away within a few months. Unfortunately, glSNe are extremely rare. Only a handful of multiply-imaged glSNe have been found to date, most of which were lensed by galaxy clusters \citep{kelly--15,rodney--21, kelly--22,chen,frye2023, 2024ApJ...967L..37P}. The first one of these, SN~Refsdal, allowed for the first time-delay measurement from glSNe, which yielded a value of $H_0 = 64.8^{+4.4}_{-4.3}$km~s$^{-1}$~Mpc$^{-1}$ \citep{Kelly_2023}. A large fraction of the error budget was linked to underlying uncertainties in modelling the complex mass structure of the galaxy cluster lens, but statistical errors remained the largest source of uncertainty (e.g. \citealt{Grillo2024}). The simpler profiles of isolated galaxy lenses leads to a smaller level of modelling uncertainty, making them ideal for precision $H_0$ measurements\cite{millon--20}. But they have shorter time delays, suffer worse microlensing and potentially larger differential dust extinction. These combine to give larger fractional time-delay measurement uncertainties (e.g. \citealt{Birrer2022, 2022A&ARv..30....8T}).

To date, two glSNe lensed by single galaxies have been discovered: iPTF16geu \citep{Goobar2017}, and SN~Zwicky \citep{Goobar2022}. Both of these glSNe, however, originated from compact systems with short time delays ($\lesssim1$ day), which did not allow for precise fractional time-delay measurements. In retrospect, this issue should have been expected for current magnitude-limited surveys as they are biased towards systems with high magnifications that tend to have shorter time-delays and smaller image separations \citep{SainzdeMurieta, SaguesCarracedo2024}. The Vera C. Rubin Observatory's Legacy Survey of Space and Time (LSST; \citealt{lsstsciencecollaboration2009lsst}) will increase the rate of galaxy-scale glSNe discoveries substantially \citep{Nikki_rates, Goldstein2019, SainzdeMurieta, Wojtak-19}, with  $\mathcal{O}(10^2)$ forecast over the survey's 10 year period. Roughly $40\%$ of discovered glSNe will have time delays that are long enough to allow for a precise ($\lesssim$ 10\% precision) time-delay measurement \citep{SainzdeMurieta}. 

Whilst strong lensing time delays are inversely proportional to $H_0$, the constant of proportionality can only be inferred from strong lens modelling. In order to obtain an accurate lens model and break degeneracies that come with lens modelling, information about a lensed host is required  \citep{Suyu2014,holicow_iv, holicow_v, holicow_xii}. In some cases, the offset between the supernova and the bright core of its host galaxy results in a glSN that does not appear to have a strongly-lensed host. In other cases, some discoverable glSNe will appear in hosts that do not reach the magnitude required to be detectable with wide-field optical surveys, and will therefore appear hostless \citep{Ryczanowski2020}. Both of these types of systems will limit our ability to obtain an accurate model of the lens and therefore to obtain a precise estimate of $H_0$.

The major challenge of the large datasets provided by Rubin will be to identify gravitationally lensed supernovae amongst the alert stream of $\mathcal{O}$($10^7$) transient alerts per night. Due to the survey's slow cadence, additional follow-up will be needed in order to obtain precise time-delay measurements \citep{Huber2019}. Several search strategies for glSNe have been suggested, e.g. \citealt{Quimby2014, Wojtak-19, Nikki_rates, Goldstein2019, Shu_lenses}. For most of these methods, the search for glSNe is hindered by our inability to distinguish them from much more common contaminants \citep{Mark_lenses}, such as un-lensed SNe hosted within potential lenses, AGN(i), faint uncatalogued variable stars, TDEs, ANTs... Monitoring known lens systems sidesteps the contaminant problem (apart from transients in the deflector), but at the cost of missing glSNe that are not in already known lens systems.

The focus of this work is to investigate the frequency of glSNe hosts being lensed and the prospects of watchlists of galaxy-scale lenses for glSNe searches. We aim to answer two questions. 1) If imaging is deep enough, what fraction of glSNe will have a multiply imaged host? 2) What fraction of glSNe will occur in galaxy-galaxy lensing systems that could plausibly be discovered beforehand in wide-field imaging surveys? To answer these questions, we calculate how many glSNe will have a strongly lensed host galaxy, and how many of these lensed hosts can be discovered with current and upcoming surveys. We use this to compute rates for glSNe that will allow for an accurate lens model. We also estimate how many more glSNe we can detect by targeting strong lenses, accounting for the rate of false positives of this method. We simulate a realistic population of gravitationally lensed supernovae and their lensed host galaxies in Section \ref{sec:simulations}, and we define our criteria for detecting our sources. In Section \ref{sec:rates} we study the rates of glSNe discoverable with this method and the properties of their systems, as well as the implications of our results for upcoming time-delay cosmography estimates. We discuss the limitations of our method and potential challenges of this strategy in Section \ref{sec:limitations} and finalise with our conclusions in Section \ref{sec:conclusions}.

\section{Simulations}
\label{sec:simulations}

\subsection{SN and host population: the SCOTCH catalogue}

We draw our source population from the Simulated Catalogue of Optical Transients and Correlated Hosts (SCOTCH; \citealt{SCOTCH}).  This simulated catalogue is based on observational host galaxy properties from GHOST \citep{GHOST}, a database of 16,175 spectroscopically classified SNe and the properties of their host galaxies that was designed to allow for realistic host-transient simulations. The transients in the SCOTCH catalogue are simulated following \cite{kessler2019} and are each associated to simulated galaxies from CosmoDC2 \citep{CosmoDC2} catalogues based on their intrinsic properties. Similar catalogues have already shown great agreement with data (e.g. \citealt{Vincenzi21}).

The SCOTCH catalogue contains 11 transient classes. For this work we simulate both SN~Ia and several subtypes of core-collapse supernovae: SN~II, SN~IIn, SN~Ib and SN~Ic. We group the last two in one class, SN~Ibc, as both of these types are hydrogen-poor “stripped-envelope” CCSNe (Modjaz et al. 2019) and are both found in hydrogen-poor environments \citep{2019NatAs...3..717M}. Following previous rate estimates \citep{Wojtak-19, Goldstein2019}, we separate the SN~IIn class, characterised by the
presence of narrow spectral line components. The remaining classes of core-collapse SNe are grouped in the SN~II class. We also simulate SLSN~I (hereafter SLSN), which are found at much lower rates than other subtypes but are extremely bright events, characterised for reaching magnitudes beyond -21 mag in optical bands \citep{2018SSRv..214...59M}.

The transients are simulated following the framework of the SNANA simulation code \citep{Kessler_2009}. The SNANA simulations include redshift-dependent volumetric rates for each transient class. As detailed in \cite{kessler2019}, these are measured from existing observations, meaning they become increasingly uncertain with redshift. The SN~Ia rate is defined as a piecewise function that follows the rates determined by \cite{Dilday2008} at redshift $z<1$ and the rate used by \cite{Hounsell2018} for redshift $z>1$, which came from measurements by \cite{Rodney2014}. The core-collapse SN rate is given by \cite{Strolger}, with the relative fractions of SN~II, SN~IIn, and SN~Ibc corresponding to \cite{Li_2011}. The rate of SLSNe appears to be consistent with the cosmic star formation history \citep{Prajs2017}. Therefore, the SLSN redshift dependent rate is modelled using the star formation history calculated by \cite{Madau_Dickinson_2014}, with $R(0) = 2\times 10^{-8}$ yr$^{-1}$Mpc$^{-3}$. As well as redshift dependent rates, the SNANA simulation code also includes spectral light curve templates for each transient class from \cite{2018PASP..130k4504P}. \cite{SCOTCH} provides details about the templates used for every transient class, which are mostly based on those presented by \cite{kessler2019}.

Each of the supernovae simulated in the SCOTCH catalogue is then assigned a host galaxy from the CosmoDC2 catalogue. In order to do this, host libraries are tailored for different transient categories containing correlations in host galaxy colour and magnitude. Due to the small sample size of some SN classes in GHOST, the hosts of archival GHOST SNe are divided into three broader categories: SN~Ia hosts, SN~II hosts, and H-poor SN hosts, which will be used for SN~Ibc and SLSNe. Using the properties of galaxies in the host libraries, a weightmap was defined, which is a multidimensional grid of host properties with values in the grid corresponding to the probability of a host with those properties to host a SN of a certain subtype. In order to match the host libraries to the observed distributions in GHOST, each galaxy was assigned a weight by interpolating the weightmap values at the value corresponding to its properties. Each SN is assigned a host galaxy based on this weight.

\begin{figure}
    \centering
    \includegraphics[width =\columnwidth]{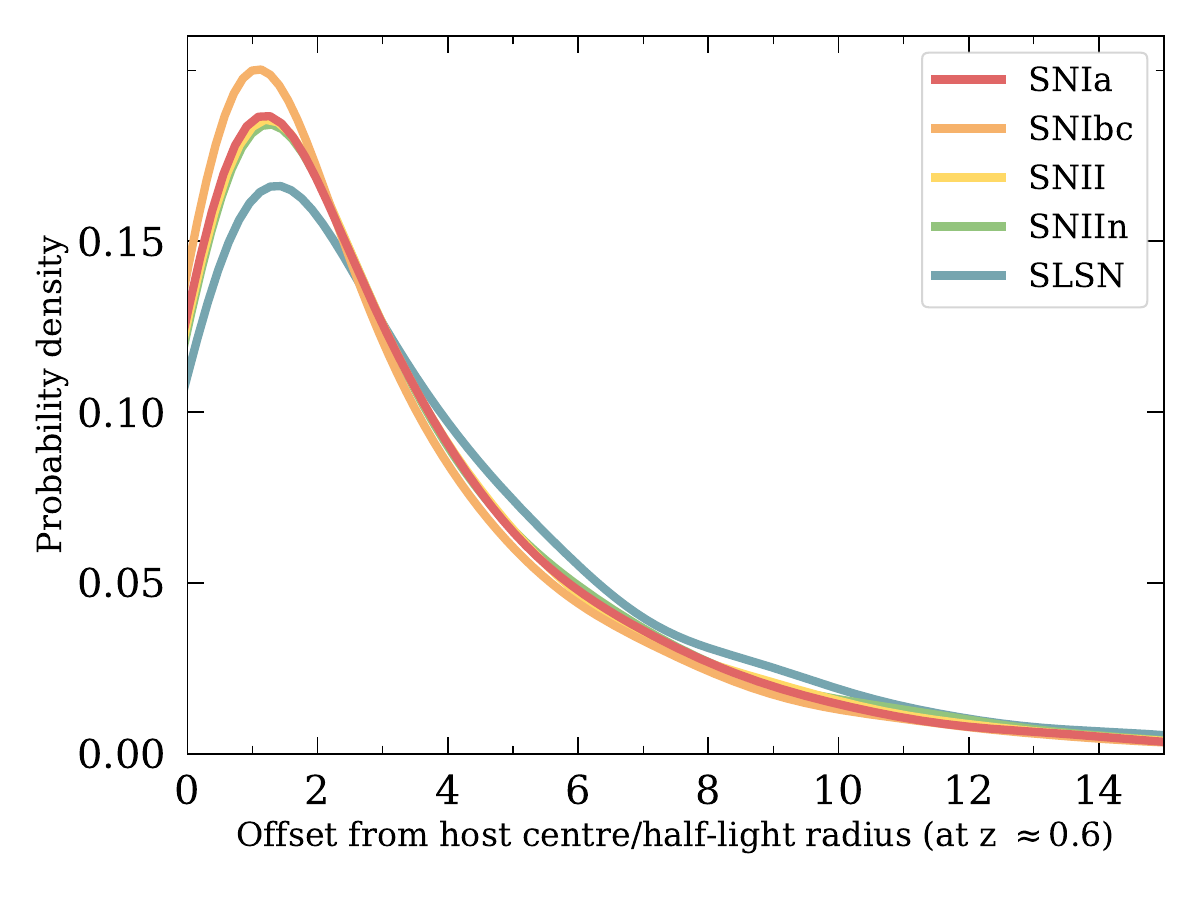}
    \caption{Relative SN offsets from the host centre in units of the host half light radius, for each source type at redshift $z = 0.6 \pm 0.05$.}
    \label{fig:offsets}
\end{figure}

After the host galaxy has been chosen, each system is assigned an offset between the supernova and the centre of the host galaxy. The distribution of these offsets is motivated by prior studies that find that SN surface density profiles roughly trace the surface brightness distribution of the disks that host them \citep{Hakobyan2009}. Despite the fact their offset prescriptions were not fitted to observations, the results were compared with SNe~Ia data from the Young Supernova Experiment (YSE) Data Release 1 \citep{YSE_DR1}, see \cite{SCOTCH} Fig. 11. We note this comparison has not been done for other subtypes, which might show statistically significant differences from our catalogue. The distribution of the SN-host galaxy offsets for a sample at redshift $z \approx 0.6$ are shown in Fig. \ref{fig:offsets}.

\subsubsection{Tidal Disruption Events and Kilonovae}

We also simulate Tidal Disruption Events (TDEs),  which occur the tidal forces of a supermassive black hole (SMBH) tear apart an orbiting star \citep{Hills1975}; and kilonovae (KNe), which are transient events due to neutron star mergers \citep{LiPazcynski1998} in order to calculate lensed transient rates. These rates are much less certain than for supernovae: following SCOTCH, we use the rates of \citep{Kochanek} and the \cite{kessler2019} templates for TDEs and the rates of \citep{Scolnic2018} (modified according to \citealt{kessler2019}) combined with the \cite{Kasen2017} templates for KNe.

We will not investigate the lensed hosts of these rare transients because SCOTCH does not include realistic host-transient associations for these. The majority of TDEs happen in the cores of the host, so if the TDE is lensed the host most likely will be. Observations of kilonovae are scarce, leading to a poor understanding of the kilonovae-host connection, especially at source redshifts relevant to lensing.

\subsection{Deflector population}
We use the methods described by \cite{Collett2015} to simulate a population of lens galaxies. Each of our lenses is assigned a density profile corresponding to a Singular Isothermal Ellipsoid (SIE; \citealt{1994A&A...284..285K}), which provides a good approximation to the mass profile of elliptical galaxies. We assume the lensing contribution of matter along the line of sight to be negligible. Thus, our simple deflector model is described by the velocity dispersion of the lens,  $\sigma_l$; the ellipticity, \textit{q}; and the lens redshift $z_l$. We use the following distributions to sample these parameters, which are the same as \cite{SainzdeMurieta}. 

The velocity dispersion distribution of galaxies is given by a modified Schechter function \citep{Sheth},
\begin{equation}
\frac{dN}{dV}=dn=\phi_{*} \left(\frac{\sigma}{\sigma_*}\right)^{\alpha}\exp\left[ - \left(\frac{\sigma}{\sigma_*}\right)^{\beta} \right]\frac{\beta}{\Gamma(\alpha/\beta)}\frac{d\sigma}{\sigma},
\label{eqn:sigma}
\end{equation}
where $\Gamma$ is the gamma function and $dn$ is the differential number of lenses per unit volume. We use the parameter values derived by \cite{bernardi} from the SDSS DR6 data describing all galaxy types, $\phi_{*}=2.099\times10^{-2}$~$(h/0.7)^3$~Mpc$^{-3}$, $\sigma_*=113.78$~km~s$^{-1}$, $\alpha=0.94$, $\beta=1.85$. We draw velocity dispersions for each of the deflectors from this distribution. We populate the sky with lenses uniformly in comoving volume and assume no redshift evolution. 

We follow \cite{Collett2015} to describe the distribution of the ellipticity for a given value of $\sigma_v$ using a Rayleigh distribution
\begin{equation}
P(1-q|s)=\frac{1-q}{s^2}\exp \Big[ - \frac{1}{2} \Big(\frac{1-q}{s}\Big)^2  \Big],
\label{eqn:prob_q}
\end{equation}
with $s=(A+B\ \sigma_v)$, A\ =\ 0.38 and B\ =\ 5.7 $\times$ $10^{-4}$ km$^{-1}$ s. This distribution takes into account that more massive galaxies tend to be closer to spherical. We truncate our distribution at $q=0.2$ to exclude highly-flattened mass profiles.

The probability of a supernova at redshift $z_s$ to have a lens with ellipticity $q$, redshift $z_l$ and velocity disperion $\sigma$ will be given by the optical depth of the system. This is defined as the area of the sky that is lensed by each lensing galaxy. It is given by
\begin{equation}
    \tau(q, zl, \sigma, z_s) = \frac{1}{A_{\mathrm{sky}}}f(q) \pi \theta_{\mathrm{Ein}}^2(z_l,\sigma, z_s),
    \label{eqn:opticaldepth}
\end{equation}
where $A_{\mathrm{sky}}$ is the area of the sky, $f(q)$ is a function that describes the ratio between the optical depth of an elliptical lens of axis ratio $q$ and a spherical lens with the same Einstein radius and $\theta_{\mathrm{Ein}}$ is the Einstein radius of the system. For an SIE, this is a function of the velocity dispersion of the galaxy, $\sigma_v$, and the angular diameter distances between observer, lens and source:
\begin{equation}
    \theta_{\rm{Ein}}^\mathrm{SIE} = 4 \pi \left(\frac{\sigma_v}{c}\right)^2 \frac{D_{\rm{ls}}}{D_{\rm{s}}}.
    \label{eqn:thetaein}
\end{equation}
Following \cite{SainzdeMurieta}, we sample lens redshifts in the range $0<z_l<1.5$, velocity dispersions in the range $100<\sigma<400$~\kms and axis ratios $0.2<q<1$ following these distributions in order to generate a mock catalogue of 100 000 lens galaxies.
\subsection{Unresolved glSNe lightcurves}
\label{sec:lightcurves}

In practice, most glSN lightcurves will appear unresolved, since typical image separations are smaller than the scatter caused by atmospheric seeing ($\lesssim$ 1") . This means that in most cases, the individual images will not be distinguishable, but will appear as a single object combining the flux from all the individual images \citep{Goldstein2019}. 

We first simulate our original unlensed lightcurves using the set of templates from \cite{kessler2019}. We aim to replicate the magnitude distribution from SCOTCH. In order to achieve this, we set our model to have the same $z$-band magnitude at peak as the lightcurves in SCOTCH once correcting for host-galaxy extinction\footnote{We draw our host $A_V$ values from the same distribution but do not have the exact value for each lightcurve. Therefore our simulated lightcurves will not be identical but will share similar properties on a population level.}.

We then apply magnification and time-delay corrections to the unlensed lightcurves, so that the combined unresolved glSN light curve is given by
\begin{equation}
    F_{\mathrm{total}}(z_s,t)=\mu_1 F_{1}(z_s,t)+\sum_{i=2}^N \mu_i F_{i}(z_s,t-\Delta t_i),
    \label{eqn:unresolved}
\end{equation}
where $F_i\left(z_s,t\right)$ is the flux derived from the spectral template, which is dependent on the SN redshift and observer-frame time, $\mu_i$ and $\Delta t_i$ are the corresponding magnifications and time delays for each of the images, and $N$ is the total number of images for each supernova.

We model the effects of reddening due to Milky Way extinction and dust in the lens and host galaxy. For Milky Way extinction we simulate our glSNe at random positions in the sky and calculate Milky Way extinction from galactic dust maps from \cite{dustmaps}. We model host galaxy extinction using the same colour law used to describe Milky Way reddening \citep{Fitzpatrick_1999}. We draw $A_V$ from a "Galactic Line of Sight" distribution, following Eqn. 2 of \cite{WoodVasey2007}, and choose $R_V$ = 3.1.  We follow \cite{feindt--19} to simulate dust at the redshift of the lens galaxies by drawing the $E(B-V)$ value from an exponential distribution with $\lambda = 0.11$.

Whilst the effects of microlensing by stars can add noise to individual glSN lightcurves \citep{DoblerKeeton2006}, it has been found to have a modest impact on population level glSNe yields \citep{Nikki_rates, SainzdeMurieta}. For simplicity, we ignore the effects of microlensing in this work.

\subsection{Defining discoverable systems}
\label{sec:discoverability}
We define different detectability criteria for time-varying and static sources. The discoverability of transient sources will be dependent on the observing strategy of the imaging survey used to detect them, while our ability to identify galaxy-scale lensing systems is highly dependent on the size of the system and our ability to distinguish lensing features (i.e arcs; \citealt{Rojas22}).
\subsubsection{Discoverable glSNe with LSST}
LSST is a 10 year optical survey of the Southern Hemisphere Sky carried out by the Vera C. Rubin Observatory \citep{2019ApJ...873..111I}. It is set to begin operations in early 2025. Roughly 90\% of its observing time will be allocated to the "Baseline Survey Strategy", a "wide-fast-deep" (WFD) survey covering about half of the sky in \textit{ugrizy} filters through 15-second exposures. The
WFD survey is expected to use a rolling cadence, which means some areas of its footprint will be assigned more frequent visits. This will improve the light curve sampling of the objects discovered in those fast-cadence areas.
In addition to this, up to $10\%$ of Rubin observing time will be dedicated to other programs, including a "deep drilling" survey that will cover a smaller area of the sky at a significantly higher cadence \citep{Bianco_2022}\footnote{Specific details about the survey observing strategy are described in  \url{https://pstn-055.lsst.io/}}.  The median effective seeing in the \textit{ugrizy} bands are $(1.10, 1.03, 0.99, 0.95, 0.93, 0.92)$ arcseconds respectively. The median single-visit 5$\sigma$ depths for the WFD fields are $(23.9, 25.0, 24.7, 24.0, 23.3, 22.1 )$ in the \textit{ugrizy} bands.

We use \texttt{simsurvey} \footnote{\url{http://github.com/ZwickyTransientFacility/simsurvey}} \citep{feindt--19} to simulate glSNe light curves as they would be observed with LSST. We use the survey plan \texttt{baseline\_v3.0} from the Operations Simulator (OpSim)\footnote{The 10-year survey plan simulations \texttt{baseline\_v3.0} can be downloaded from \url{http://astro-lsst-01.astro.washington.edu:8080/}}. This observing strategy uses two filters at a time using a mix of uniform cadence and a half-sky rolling cadence in the "deep drilling" survey region. Updated versions of this strategy do not introduce significant changes that would impact our strategy \citep{Nikki_rates}.
For the CCDs, we rescale the ZTF corners to account for the field of view of LSST. We define different discoverability criteria: one 3$\sigma$ detection, one 5$\sigma$ detection, and three 5$\sigma$ detections (all in any of the bands) to explore the potential of this method to identify glSNe earlier in order to request follow-up observations. The glSNe are distributed evenly across the LSST footprint. We simulate 10,000 glSNe per redshift bin of size $\Delta z = 0.1$ and their lightcurves are simulated as described in Section \ref{sec:lightcurves}.

\subsubsection{Discoverable lenses}
A fraction of the glSNe will also have a magnified host; however, not all of them will be detectable as strong lenses, either because the host will be fainter than the magnitude limit of the survey, or due to lack of obvious shearing. Most imaging-based strong lens searches require at least some level of human input. We establish a set of requirements for a lens to be classified as detectable. Firstly, and most importantly, we need the centre of the host galaxy to be magnified, that is
\begin{equation}
    \theta_{\mathrm{Ein}}^2>x^2_s + y^2_s,
    \label{eqn: lensed host}
\end{equation}
where $\theta_{\mathrm{Ein}}$ is the Einstein radius of the system and $(x_s, y_s)$ are the unlensed source positions. Human lens finders will look for typically arc-like features, the presence of a similar-colour counter image, bluer features, or a morphology typical of a strongly lensed system. We summarise this by requiring tangential shearing of the arcs to be observable, so
\begin{equation}
    \mu_{\mathrm{TOT}}>3,
\end{equation}
where $\mu_{\mathrm{TOT}}$ is the total magnification of the source galaxy. We also require there to be enough separation between the features in order for us to distinguish between the different images
\begin{equation}
    \theta_{\mathrm{Ein}}>\mathrm{seeing}.
    \label{eqn:seeing}
\end{equation}
Finally, we require the lensed galaxy to be brighter than some magnitude threshold $m_{\mathrm{thres}}$ defined based on the survey,
\begin{equation}
    m_\mathrm{host}-2.5\mathrm{log}(\mu_{\mathrm{TOT}})<m_{\mathrm{thres}}.
\end{equation}
Due to these requirements, we consider two types of surveys. Upcoming ground based surveys, such as LSST, will show an improvement in limiting magnitude, but the seeing is impacted by atmospheric effects. Space-based surveys such as Euclid avoid this issue, leading to better image resolution. These surveys have the ability to distinguish lenses with smaller Einstein radii (Eqn.~\ref{eqn:seeing}), but will require brighter lenses and will survey a smaller fraction of the sky. For these reasons, we set the LSST $i$-band threshold to 25~magnitudes and the seeing cut at $0.8$ arcseconds. For Euclid, we set the magnitude threshold at $m_{\mathrm{thres}} = 22.5$ magnitudes in the $i$-band and a seeing of 0.4 arcseconds.

The synergy between these surveys has already been suggested for supernova studies, as combining data in the Euclid Deep Drilling Fields with the LSST deep field observations can reduce systematics in measuring SN distances at higher redshifts \citep{Euclid_LSST}. In our case we aim to maximize our chances of identifying gravitational lenses by combining the strengths of both survey types: increased sensitivity to faint lensing events and enhanced resolution for resolving lensing configurations in more compact systems.

\subsection{Sample weighting}
\label{sec:weights}
 To calculate rate estimates of glSNe for each class of SN, we calculate the optical depth for each lensing system using Eqn. \ref{eqn:opticaldepth}. Each glSN will then be assigned a weight
 \begin{equation}
     w_i = \mathrm{f}_{\mathrm{sky}}\frac{N_{\mathrm{glSNe}}}{N_{\mathrm{sim}}}\tau(q, z_l, \sigma, z_s)p_{\mathrm{det, i}} 
 \end{equation}
where $\mathrm{f}_{\mathrm{sky}}$ is the fraction of the sky we simulated using \texttt{simsurvey} (which in our case is 75\% of the sky), $N_{\mathrm{sim}}$ is the size of our catalogue, $\tau$ is the optical depth and $p_{\mathrm{det, i}}$ is the number of times the supernova was recovered with \texttt{simsurvey} using the LSST observing strategy divided by the number of times it was simulated. The total number of glSNe in the sky for each subtype, $N_\mathrm{glSNe}$ is calculated by computing the integral
\begin{equation}
    N_{\mathrm{glSN}} = \int_0^2 dz_s \frac{1}{1+z_s}\frac{dN}{dz_s}P_{\mathrm{SL}}(z_s),
    \label{eqn:N_glSNe}
\end{equation}
where $dN$/$dz_s$ is the volumetric rate of glSNe of a certain type and the probability of a supernova being strongly lensed, $P_{\mathrm{SL}}(z_s)$ is given by
\begin{equation}
   P_{\mathrm{SL}}(z_s) =  \int_0^{1.5}dz_l \int_{0.2}^1dq  \int_{100}^{400} d\sigma \tau(q, z_l, \sigma, z_s) p(q|\sigma) \frac{dN}{dz_ld\sigma}.
\end{equation}
 The probability $p(q|\sigma)$ represents the distribution of axis ratio given a velocity dispersion, and is derived from Eqn. \ref{eqn:prob_q}. We calculate ${dN}/{dz_ld\sigma}$ from combining Eqn. \ref{eqn:sigma} with the differential comoving volume. 
The number of discovered glSNe of each subtype will then be given by the sum of the weights
\begin{equation}
    \sum_i w_i = N_{\mathrm{glSNe,det}} / \ \mathrm{year}.
\end{equation}

\begin{figure*}
    \centering
    \includegraphics[width=\textwidth]{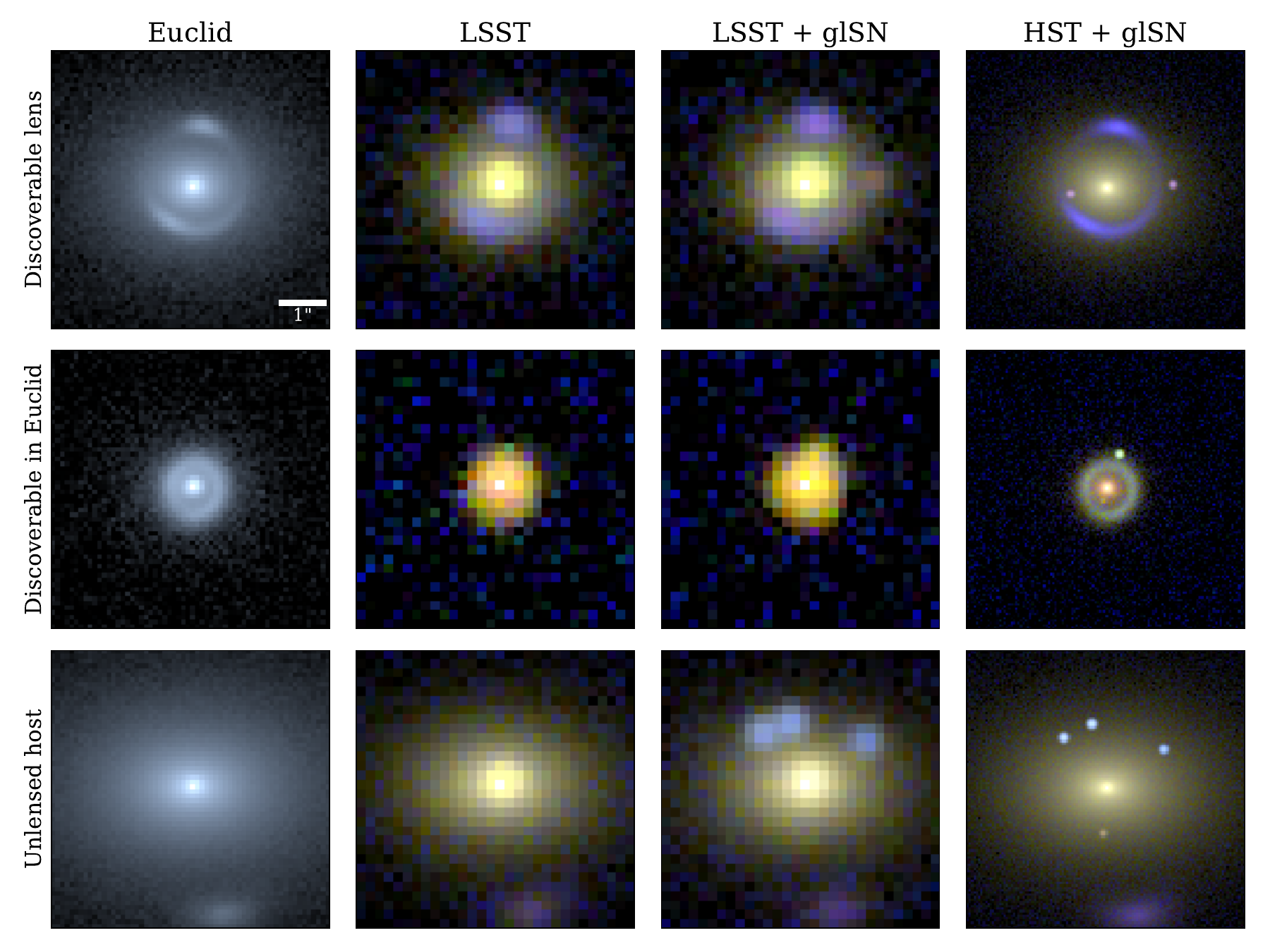}
    \caption{Sample of simulated lenses from our catalogue. The first two columns show the lenses without the supernova. Column 1 shows a mock Euclid $VIS$-band image and Column 2 a mock LSST $gri$ composite. Column 3 shows a mock LSST image including the supernova images at peak luminosity of the first image. Column 4 shows what the system would look like with an orbit of HST follow up ($gri$, with 960s per band), including the supernova images. The lens in Row 1 would be discoverable in both LSST and Euclid. Row 2 shows a lens that would be discoverable in Euclid but not LSST and Row 3 shows an example of a glSN without a lensed host, similar to SN~Zwicky.}
    \label{fig:mock_lenses}
\end{figure*}

To perform our simulations, we sample 10 000 SNe and hosts from the SCOTCH catalogues for each SN subtype. Each of these SNe are then assigned 10 lens galaxies from our deflector catalogue, drawn from a distribution proportional to the optical depth of the lensing systems. Each SN is placed within the area of strong lensing of the lens. We then use \texttt{lenstronomy} \footnote{\url{ https://github.com/sibirrer/lenstronomy}}  \citep{BIRRER2018189,2021JOSS....6.3283B} to compute the positions of their images, as well as the magnification and time-delays for each of the images. We use the offset between the SN and its host galaxy given in the SCOTCH catalogues to define the host galaxy position, and use \texttt{LensPop}\footnote{\url{ https://github.com/tcollett/LensPop}} \citep{Collett2015} to solve the lens equation and simulate images for the galaxies in our catalogue. A sample of these simulated images can be found in Fig. \ref{fig:mock_lenses}. If the host is strongly lensed, we recover its total magnification and magnitude in different bands.

\section{Results and Discussion}
\label{sec:rates}
\subsection{Rates of lensed transients}

We follow the method described in Section \ref{sec:weights} to calculate the annual rates of different SN subtypes discovered by LSST. These are shown in Table \ref{tab:rates}. We find a total of 88 glSNe per year will be discoverable with LSST, out of which 39 will be glSNe~Ia, 6 will be glSNe~Ibc, 41 will be of Type II, 1.5  will be Type IIn and 0.5 will be SLSNe. We also forecast 0.2 lensed TDEs and 10$^{-3}$ lensed kilonovae per year. These rates are sensitive to a range of assumptions, and their uncertainties are discussed in detail in Section \ref{sec:limitations}. The discoverable glSNe are found in systems with median velocity dispersion $\sigma_v = 197$~km~s$^{-1}$, lens redshift $z_l=0.28$, source redshift $z_s=0.68$, and Einstein radius $\theta_{\mathrm{Ein}}=0.53''$. The median longest\footnote{quads have more than one-time delay} time-delay in these systems is $\Delta t_{\mathrm{max}}$ = 16 days, and the median maximum angular separation between images of the same SN is $\theta_{\mathrm{max}}=1.02''$. This means a large fraction of glSNe will have at least partially resolved images and time-delays that are long enough to allow for precise time-delay measurements ($\Delta$t>10 days; \citealt{Pierel2019}).

\begin{table}
\centering
\begin{tabular}{|c|c|}
\hline
\textbf{Class} & \textbf{Lensed Rate (yr$^{-1}$)} \\
\hline
SN~Ia & 39 \\
SN~Ibc & 6 \\
SN~II & 41 \\
SN~IIn & 1.5\\
SLSN & 0.5\\ 
TDE & 0.2\\ 
Kilonovae & $10^{-3}$\\ 
\hline

\textbf{Total} & \textbf{$\approx$ 88}\\
\hline
\end{tabular}
\caption{Yearly rates of lensed transients by class}
\label{tab:rates}
\end{table}

\subsection{Lensed hosts of glSNe in LSST and Euclid}
\label{sec:lensproperties}
\begin{figure}
    \centering
    \includegraphics[width=\columnwidth]{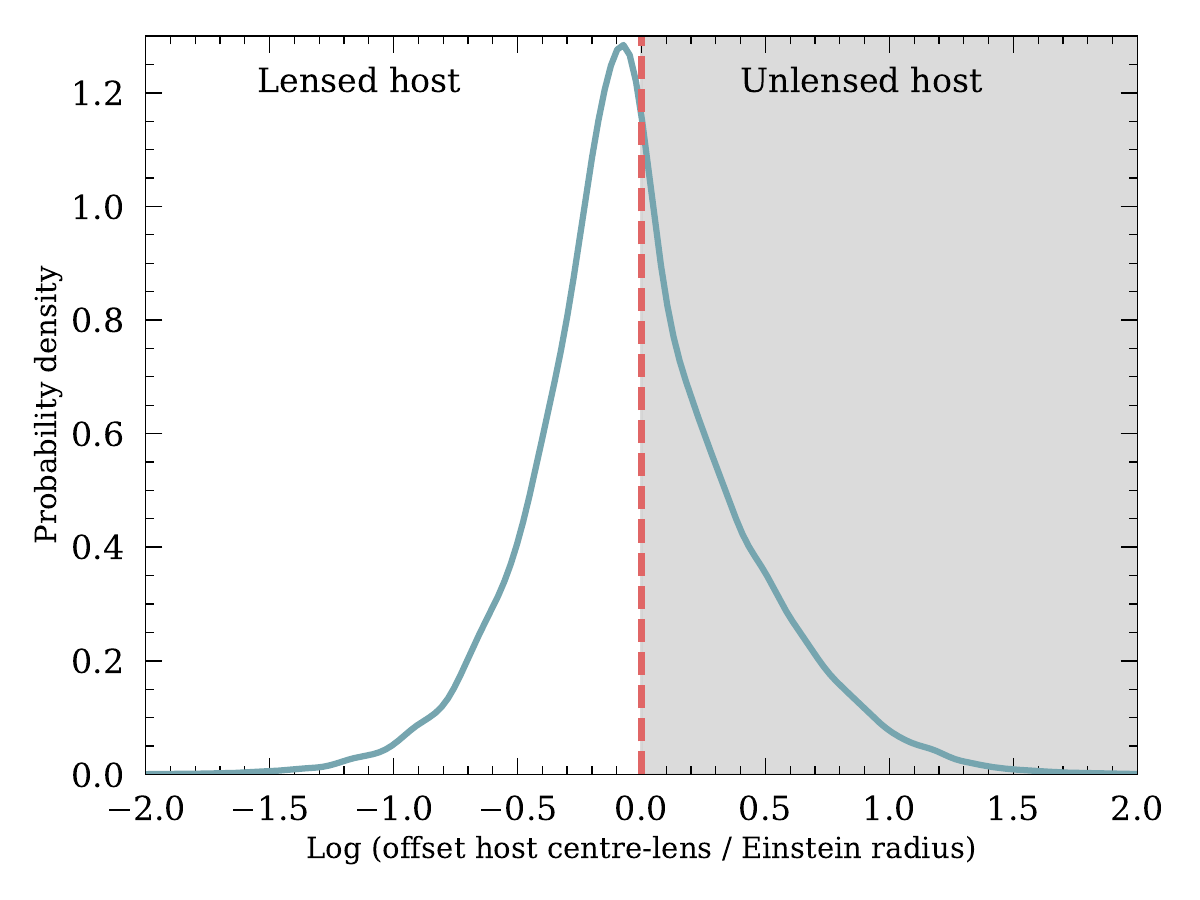}
    \caption{Offset between the centre of the lens and the centre of the glSN host in units of the Einstein radius of the system. 54\% of discoverable glSN host centres will be found in the unshaded region, and therefore will be strongly lensed.}
    \label{fig:galaxies}
\end{figure}

With our calculated rates of discovered glSNe, we now investigate which systems will also feature strongly lensed host galaxies. Figure \ref{fig:galaxies} shows the distribution of offsets between the centre of the host galaxy and the centre of the lens. We find that $\approx 54$ \% of glSNe discoverable with LSST will have a host that is located within the area of strong lensing. Dividing this result by subtype, we find 51\% of glSNe~Ia will have a strongly lensed host, 42\% of glSNe~Ibc, 58\% of glSNe~II, 46\% of glSNe~IIn and 68\% of glSLSNe. Only 33\% of galaxies have a total magnification $\mu_{\mathrm{TOT,host}}>3$, required for observing tangential shearing of the arcs as explained in Section \ref{sec:discoverability}. Table \ref{tab:lensedhosts} shows a breakdown by SN subtype of these results.

\begin{table}
\centering
\begin{tabular}{|c|r|r|}
\hline
Class & $\mu_{\mathrm{TOT,host}} >2$ & $\mu_{\mathrm{TOT,host}}>3$ \\
\hline
SN~Ia & 51\% & 30\%\\
SN~Ibc & 42\% & 26\% \\
SN~II & 58\% & 36\%\\
SN~IIn & 46\% & 28\%\\
SLSN & 68\% & 37\% \\ \hline
\textbf{Total} & \textbf{54\%}  & \textbf{33\%}\\
\hline
\end{tabular}
\caption{Fraction of glSNe with a strongly lensed host per subtype. The total fraction is the average weighted by the relative rates of each subtype.}
\label{tab:lensedhosts}
\end{table}

In reality, not all of these hosts will be discoverable with current and upcoming surveys. As detailed in \cite{Ryczanowski2020}, some of these will be fainter than the detection threshold of the survey, making the glSNe appear hostless. In order to define what fraction of these will be detectable with LSST and Euclid, we need to specify our discovery criteria for each of these surveys.

In Section \ref{sec:discoverability} we discussed the conditions for discovering a lensed host. For the case of LSST, we defined a magnitude threshold requiring objects to be brighter than 25~mag in the $i$-band and an Einstein radius threshold $\theta_{\mathrm{Ein}}>0.8"$. We find 40\% of glSNe host galaxies have these characteristics. For the Euclid survey, we require objects to have a VIS magnitude\footnote{$\mathrm{VIS} = \frac{1}{3}(r+i+z)$, following \cite{Collett2015}} brighter than 22.5 mag and  Einstein radius $\theta_{\mathrm{Ein}}>0.4"$. We find that 68\% of the systems have these characteristics. Between the two surveys, a total of 72\% glSN hosts will be discoverable as lensing systems. Table \ref{tab:percentages} shows these results for each SN subtype. 

We note 25 \% of lensed hosts are not discoverable with either of these surveys. Upcoming deeper space-based surveys such as the Nancy Grace Roman Space Telescope (Roman; \citealt{Spergel2015}) will be able to push the limiting magnitude threshold required for detection, increasing the fraction of discoverable glSN hosts and rates of glSNe \citep{2021ApJ...908..190P}. The combination of an Einstein radius threshold of $\theta_{\mathrm{Ein}}>0.4"$ and a magnitude threshold of 25 mag., merging both of the advantages of LSST and Euclid, would allow for the discovery of 82\% of lensed hosts. This means 18\% of systems will still be too compact to be detectable, but these systems are less likely to be interesting for time-delay cosmography and would pose their own challenges for lens modelling.

It is worth noting that the rates cited require ideal lens discovery conditions. Whilst substantial progress has been made in reducing billion object surveys to samples of 1000s of lens candidates \citep{More2016, Jacobs2019, KiDS2021, Rojas22} scaling strong lens discovery systems up to Euclid and LSST has not yet been demonstrated. We examine how much our values change for more pessimistic discovery prospects. For the case of LSST, if we adjust the minimum Einstein radius threshold to $\theta_{\mathrm{Ein}}>1"$, we find 25\% of lensed glSN hosts are now discoverable, as opposed to our previous 40 \%. When we increase the Einstein radius threshold for our Euclid lenses to $\theta_{\mathrm{Ein}}>0.5"$, the loss is less severe, with 59\% of the strongly lensed hosts still being discoverable (c.f. 68\% before). Alternatively, we can consider the case in which we are able to meet the Einstein radius discovery threshold but do not reach the discovery depth. For the case of LSST, reducing the magnitude threshold to 24 still yields $\approx 40\%$ of lensed hosts, with a negligible decrease in sample size. Euclid rates are much more sensitive to a loss of depth: reducing the VIS threshold depth by a magnitude (to 21.5) loses a fifth of the Euclid discoverable lens sample, with 52\% of all lensed hosts now being discoverable in Euclid data alone. 

In the worst case scenario, with the more pessimistic thresholds in for both Einstein radii and limiting magnitude, we find $54\%$ of lensed hosts are discoverable, with $25\%$ coming from LSST and $46\%$ from Euclid, and an overlap of 17\% between both surveys. These results highlight that many of the lensed SNe host systems will be close to detection thresholds as galaxy-galaxy lenses. Combining surveys that reach fainter depths and better angular resolution for galaxy-galaxy lens discovery will be important to mitigate risk and maximise the discovery potential for galaxy-galaxy lensing.

\begin{table*}
\centering
\begin{tabular}{l|c|ccccc}
\hline
\textbf{}                                    & \textbf{Total}   & \textbf{SN Ia}  & \textbf{SN Ibc} & \textbf{SN II} & \textbf{SNIIn}  & \textbf{SLSN}   \\ \hline
\textbf{Discoverable lenses (LSST + Euclid)} & \textbf{75\%}  & \textbf{72\%}         & \textbf{81\%}          & \textbf{76\%}  & \textbf{73\%}         & \textbf{43\%}          \\ 
LSST                                         & \textbf{40\%}  & 38\%          & 44\%          & 42\%          & 39\%   & 30\%         \\ 
Euclid                                       & \textbf{68\%}  & 66\%          & 79\%          & 69\%          & 67\% & 23\%          \\ 
Both                                         & \textbf{34\%}  & 32\%          & 43\%          & 35\%          & 33\% & 10\%          \\ \hline
\textbf{Total fraction with a lensed discoverable host} &
  \multicolumn{1}{c|}{\textbf{25\%}} &
  \multicolumn{1}{c}{\textbf{22\%}} &
  \multicolumn{1}{c}{\textbf{21\%}} &
  \multicolumn{1}{c}{\textbf{27\%}} &
  \multicolumn{1}{c}{\textbf{20\%}} &
  \multicolumn{1}{c}{\textbf{16\%}} \\ 

\end{tabular}%
\caption{Fraction of lensed hosts with $\mu_{\mathrm{TOT}}>3$ discoverable with different imaging surveys}
\label{tab:percentages}

\end{table*}

\begin{figure*}
    \centering
    \includegraphics[width=\textwidth]{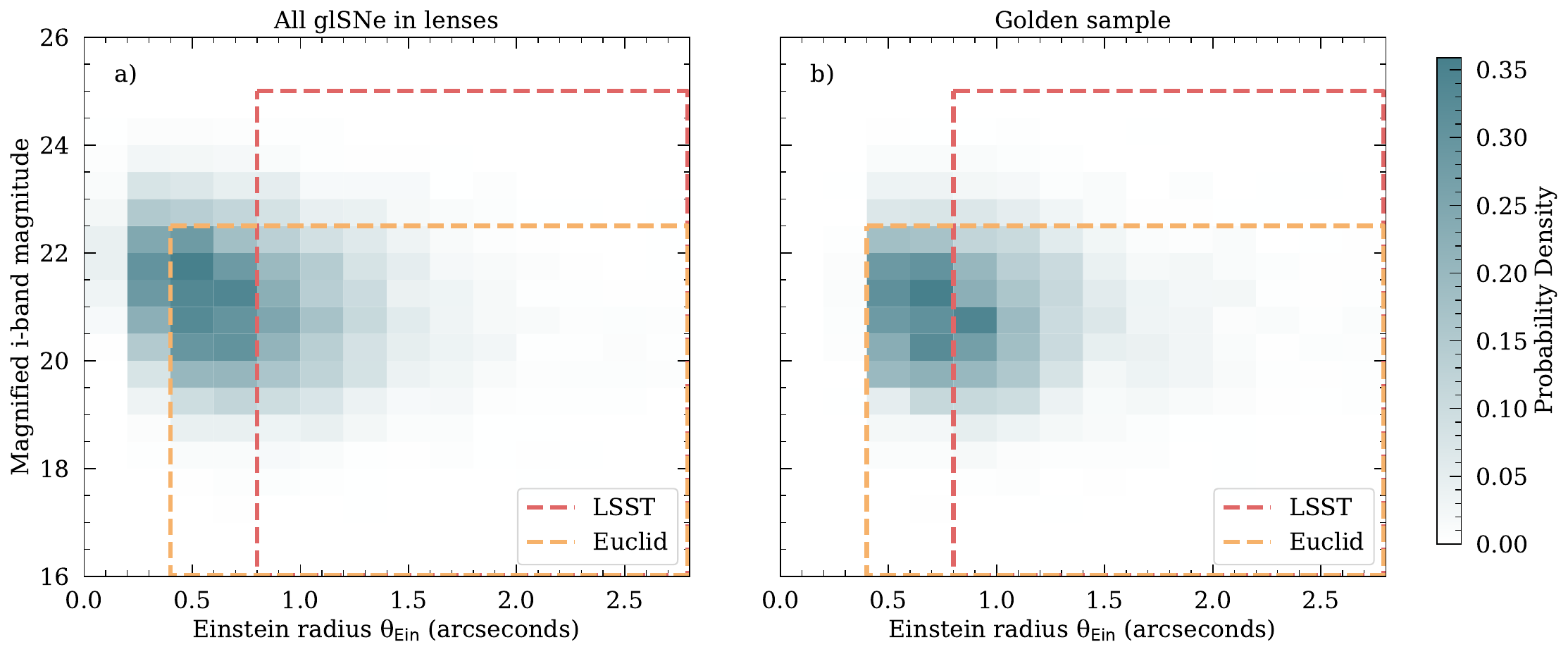}
    \caption{Distribution of detectable lensed host galaxies as a function of source i-band magnitude and Einstein radius. The nominal discovery regions of LSST and Euclid are shown in red and orange respectively. The left panel shows all glSNe host systems whereas the right panel only shows golden glSNe for cosmology with long time delays and large image separations. Since the golden sample is biased towards larger image separations a larger fraction have a discoverable host.}
    \label{fig:LSST lenses}
\end{figure*}

\begin{figure*}
 \centering
    \includegraphics[width=\textwidth]{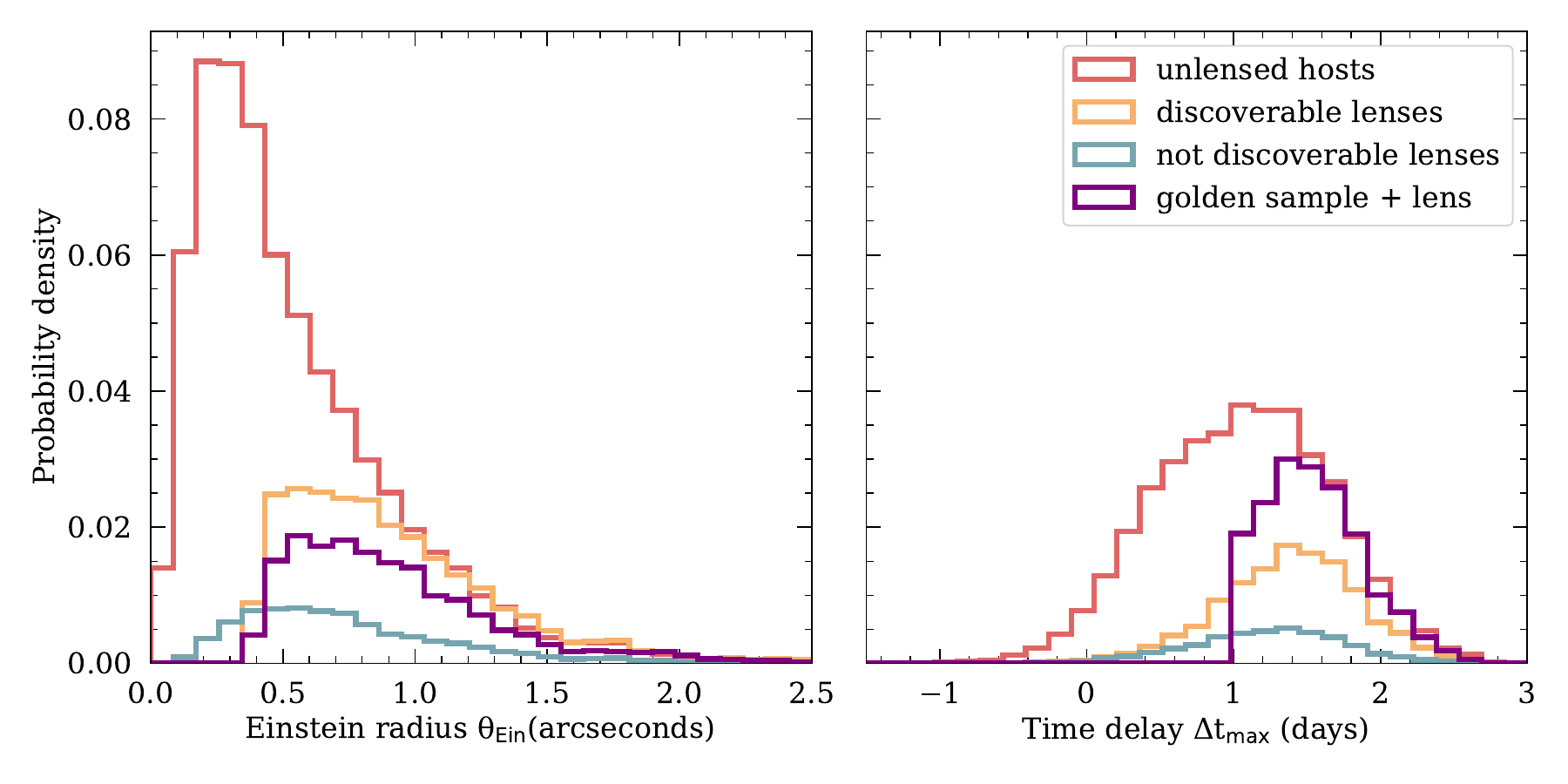}
    \caption{Properties of lensed SN systems, broken down into subpopulations. Red shows glSNe where the host is not multiply imaged. For Yellow, the host is multiply imaged and discoverable as a galaxy-galaxy lens in either Euclid or LSST. Blue shows systems where the host is multiply imaged but not identifiable as such with LSST or Euclid. Purple are the glSNe with lensed hosts, time delay > 10 days and image separation > 0.8 arcseconds. Overall, lensed systems have greater Einstein radii and longer time-delays.}
    \label{fig:discoverable_properties}   
\end{figure*}

\begin{figure}
 \centering
    \includegraphics[width=\columnwidth]{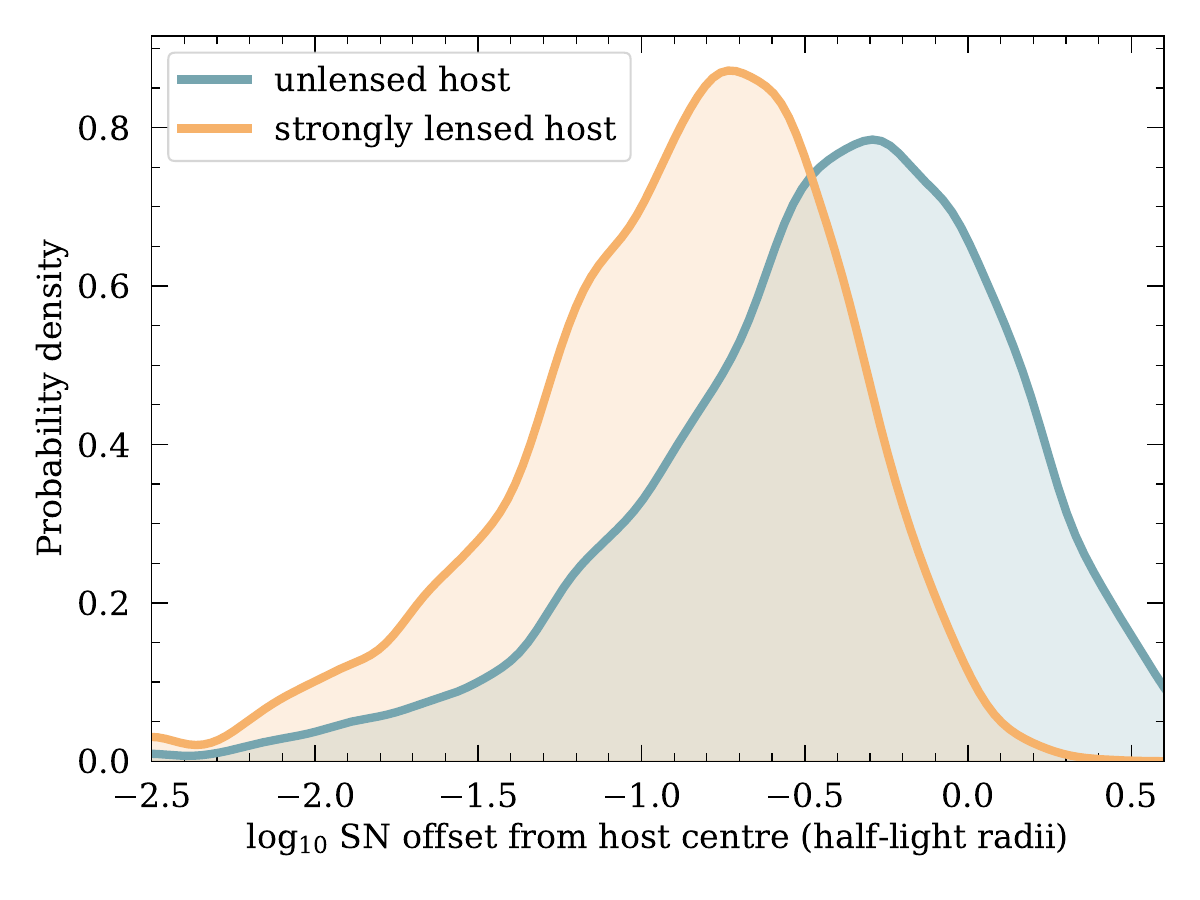}
    \caption{SN offset from its host centre in units of host half-light radii. SNe in strongly lensed hosts are located closer to the nucleus of their host galaxy than those in unlensed hosts.}
    \label{fig:discoverable_properties2}   
\end{figure}

We also consider whether the properties of the discoverable lenses differ from those that are not discoverable, and explore how both of these populations differ from the population of systems with unlensed hosts, shown in Fig. \ref{fig:discoverable_properties}. As expected, the main differences between the population of discoverable lensed hosts and not discoverable lensed hosts are driven by our angular separation requirement. The median Einstein radius of discoverable lenses is $\theta_{\mathrm{Ein}}\approx 0.84"$, whereas the median Einstein radius of not discoverable lenses is $\theta_{\mathrm{Ein}}\approx 0.67"$. The median Einstein radius of unlensed systems peaks at lower Einstein radii than the original population of discoverable glSNe, with $\theta_{\mathrm{Ein}}\approx 0.44"$. The median time-delay for unlensed hosts is 13 days, whereas for discoverable lenses is 29 days and for not discoverable lenses it is 23 days. The redshift distribution of sources peaks at $z_s\approx 0.69$ for discoverable sources and $z_s\approx 0.72$ for not discoverable lenses, as opposed to  $z_s\approx 0.63$ for unlensed hosts. The properties of the lenses in our sample are once again determined by our Einstein radius cut, as slightly higher velocity dispersions are preferred for discoverable systems, with mean $\sigma_v = 229$ km~s$^{-1}$. Not discoverable lenses have $\sigma_v = 211$ km~s$^{-1}$ and unlensed hosts have a mean $\sigma_v = 187$~\kms. These results show that the Einstein radius cut for discoverability has a significant impact on the glSNe properties. 
Figure \ref{fig:discoverable_properties2} shows the distribution of glSN offsets from their host galaxy centre for strongly lensed hosts and unlensed hosts. We show that glSNe in galaxy-galaxy lensing systems happen closer to the centre of their host galaxy, with a median separation of 0.19 half-light radii whereas those in unlensed hosts have a median separation of 0.35 half-light radii. This is expected, as the closer the SN is to the host galaxy, the more likely it is for the host to also be strongly lensed. For very compact systems where the supernova happens close to its host, this will pose the additional challenge of being able to resolve the supernova from the nucleus of the host galaxy.

\subsection{Implications for precision cosmology}
\label{sec:cosmography}
A fraction of discoverable glSNe will belong to the "golden sample" for time-delay cosmography, which is defined as the systems for which we believe we can confidently measure a precise ($\sim$10\% precision) time-delay measurement. We define this as systems with image separations of $\theta_{\mathrm{max}}$> 0.8'' and $\Delta t_{\mathrm{max}}$ >10 days, due to the fact that we can measure time-delays at the $\approx 1$d level (\citealt{Pierel2019}). So far these requirements have been simply defined for glSNe~Ia, partly due to their standardisable nature and also due to the fact they are the only subtype that has been detected so far in galaxy-scale lenses. For this reason, in this subsection we will focus on glSNe~Ia. It is worth noting that with the increased detection threshold of LSST, more glSNe of all subtypes will be observable, and time-delay measurements will be obtainable from other SN subtypes, although our current ability to measure time-delays from other subtypes will most likely require longer time delays \citep{2022A&A...658A.157H, 2021ApJ...908..190P}. 

 As we have shown in Section \ref{sec:lensproperties}, glSNe in strongly lensed systems have slightly greater angular separations and time-delays than those systems in which the host is not strongly lensed.  This is because of a selection bias: the physical separation between host and SN is independent of the lens, but the larger the Einstein radius of the lens the greater the chance that both host and SN are multiply imaged. Time delays and image separations scale with the Einstein radius.
 
 40\% of glSNe~Ia in the golden sample have a lensed host with $\mu_{\mathrm{TOT, host}}>3$, compared to $30\%$ of the whole sample, and the majority of these (91\%) systems will be discoverable with either LSST or Euclid. This is because the golden sample requires long time delays and image separation which implies larger than typical Einstein radii. Fig. \ref{fig:discoverable_properties}b shows the Einstein radius of lenses and lensed source magnitudes for the golden sample. 
 
 47\% of glSNe~Ia in our simulations, that is $\approx 18$ per year, belong to this "golden sample". According to this $\approx 7 $ glSNe~Ia per year will have the properties suitable for time-delay cosmography and occur in a system that could already have been identified as a galaxy-galaxy lens. Over the 10 years of the survey, this will add up to 70 glSNe and assume a $10\%$ error $H_0$ measurement per system, this would imply an $\approx 1.3\%$ precise measurement of the Hubble constant from glSNe detected through monitoring galaxy-galaxy lenses.

\subsection{Time of discovery and follow-up}

Normally, multiple high signal-to-noise ratio (S/N) observations are required before claiming that a new transient could be lensed. When monitoring known lensing systems, if a new transient is discovered we can immediately classify the transient as interesting since the chance of it being lensed is much higher than for a random field. This would enable much faster spectroscopic confirmation and follow-up observations to begin.

\begin{figure}
    \centering
    \includegraphics[width=\columnwidth]{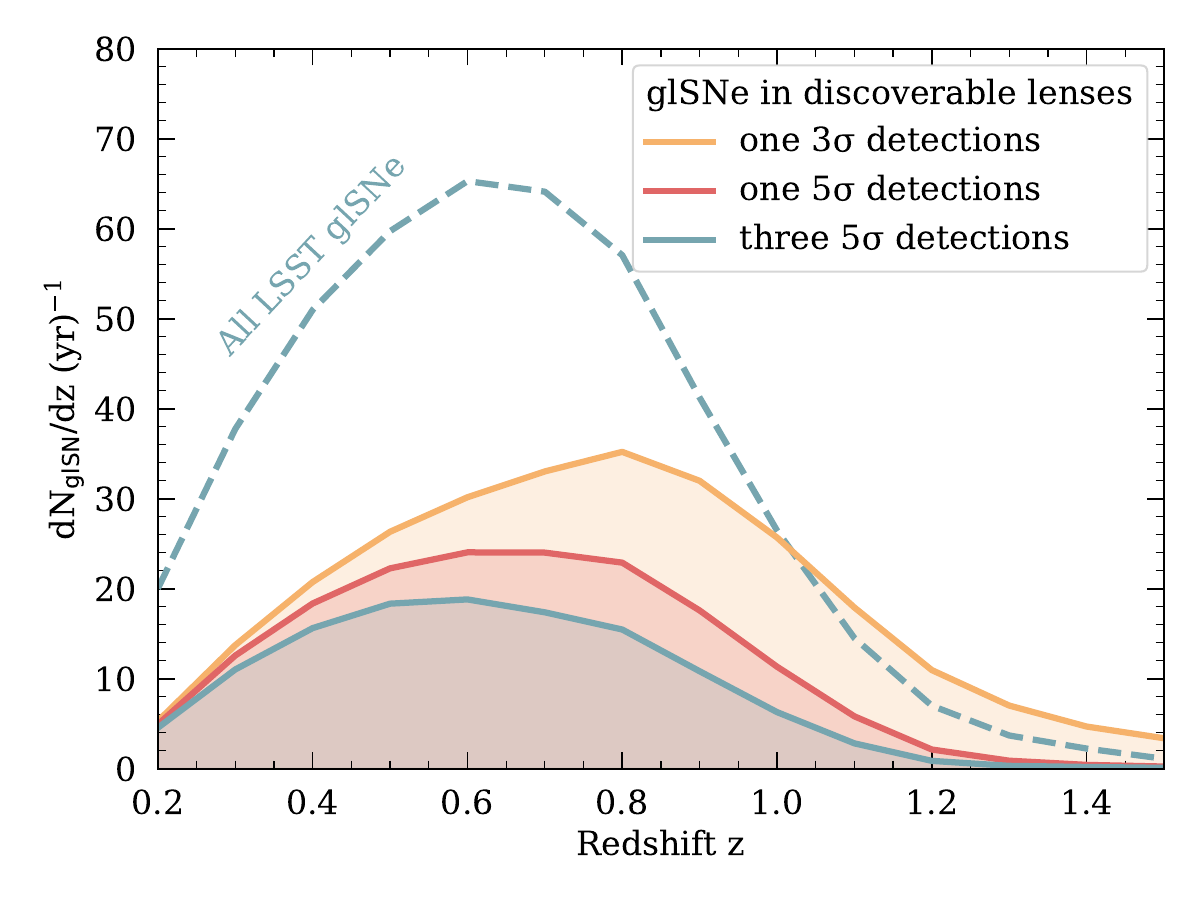}
    \caption{Rate of glSNe~Ia discovered in LSST for different discovery detection requirements. Reducing the S/N requirement substantially increases the yield of glSNe past $z$=1}
    \label{fig:simsurvey1}
\end{figure}

In this section, we explore how many more glSNe can be detected in discoverable lenses if we reduce the number of detections required for "discovery". We compare the results from our initial analysis, which requires three 5$\sigma$ observations, with the more relaxed requirements of one 5$\sigma$ and one 3$\sigma$ observation. We first look at the change in the fraction of simulated systems that were detected as a function of the glSN redshift for each of our detection cuts, shown in Fig. \ref{fig:simsurvey1}. Relaxing our detection cuts from three to just one 5$\sigma$ detection allows for an 8\% increase in rates. Setting our detection requirement to a single 3$\sigma$ detection not only increases rates by 30\% with respect to our initial cuts, but also increases the maximum redshift at which glSNe can be detected. This is because normally higher redshift observations have lower S/N.

The success of time-delay cosmography also relies on our ability to detect transients as soon as possible in order to request follow-up observations, as other studies have shown that LSST alone does not have the cadence required for a good enough sampling of light curves to do precision time-delay measurements without follow-up from other telescopes \citep{Huber2019}. We first consider at what time in the observer-frame glSNe can be detected. We find 29\% glSNe~Ia are discovered 10 days before their B-band peak. Requiring just one detection this fraction becomes 42\%, and 45\% for a 3$\sigma$ detection. Ideally, we would detect our glSNe 20 observer-frame days before peak in order to maximise the chance of obtaining follow-up observations. For the three 5$\sigma$ requirement only 4\% ($\approx 2$ per year) satisfy this requirement. For the one 5$\sigma$ requirement this fraction increases to 8\%, and further increases to 15\% for the one 3$\sigma$ requirement. This is driven by the fact the 3$\sigma$ detection allows us to detect glSNe at higher redshift, which have a slower evolution as their lightcurve is stretched due to time dilation.

\begin{figure}
    \centering
    \includegraphics[width=\columnwidth]{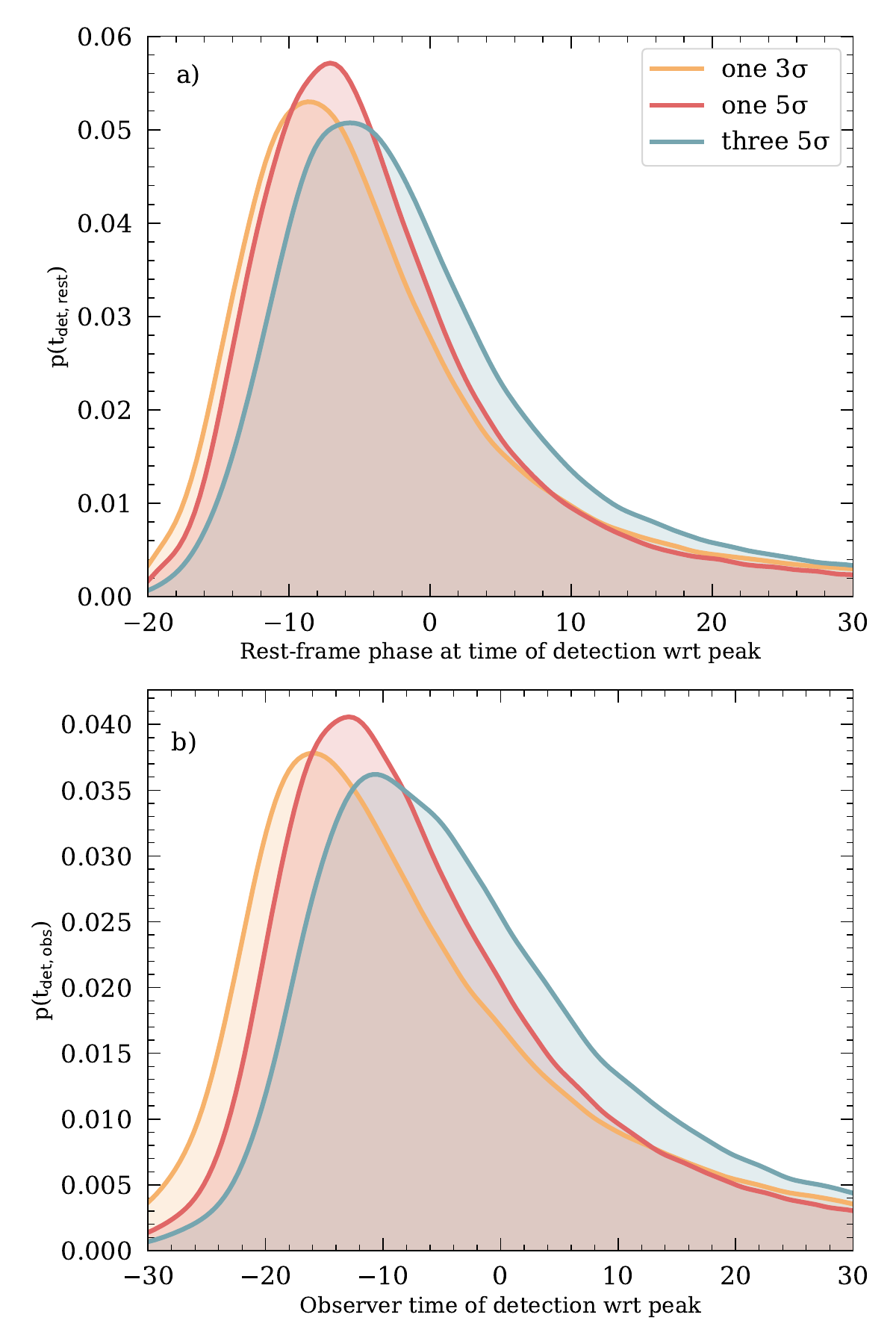}
    \caption{Probability density function for the time of discovery of of glSNe~Ia in LSST. The top (bottom) panel shows time relative to peak in the rest-frame (observer-frame). }
    \label{fig:simsurvey2}
\end{figure}

 We convert these times to rest-frame to find at what time in their evolution glSNe will be detected. The distribution of discovery phases of the first images occurring in discoverable lenses is shown in Fig. \ref{fig:simsurvey2}. For our 3 5$\sigma$ detection requirement, 14 \% of glSNe~Ia are discovered 10 days before their B-band peak. Requiring just one 5$\sigma$ detection, this fraction becomes 22\% and 26\% for a single 3$\sigma$ detection.

\subsubsection{False positives - the rate of unlensed SNe in lensing galaxies}
A lower S/N requirement could be beneficial for finding more glSNe, but also comes at the cost of a higher false positive rate. When a transient is discovered in a known lensing system it could be a glSN, but it could also be an unlensed supernova in the lens galaxy. With a significant number of observations in different colour bands, this second scenario can be discarded by fitting observations to supernova templates at the redshift of the lens. Robustly distinguishing this scenario however requires a considerable number of observations to compare against the SN template. It cannot be done with a single detection. 

In order to estimate the rate of false positives, we first calculate the observer-frame SN rate in the foreground galaxies. In order to do this we follow the method described by \cite{2012JCAP...11..015L} to obtain the rest-frame SN~Ia rate for each lens galaxy, which depends on both the stellar mass and star-formation rate of the galaxy.

The stellar mass for each of the lenses is determined from their velocity dispersion following \cite{stellarmass}. We choose the expression for the velocity dispersion ($\sigma_v$)-stellar mass($M_{\star}$) relation to be
\begin{equation}
\mathrm{log}\left(\frac{\sigma_v}{\mathrm{km}  \mathrm{s^{-1}}}\right) \approx 2.21 + 0.18 \ \mathrm{log}\left(\frac{M_\star}{10^{11} M_\odot}\right) + 0.4 \ \mathrm{log}(1+z).
\end{equation}
In their work, this expression is valid for a fiducial sample with redshift $z<1$. We choose this due to the fact most of our lenses will be found up to redshift $z$ $\approx$ 1 and LSST will not be able to discover unlensed SNe~Ia beyond that redshift.
As expected, our lenses are massive galaxies, with a median stellar mass log $M_{\star}/M_{\odot}$ = $11.2 ^ {+ 0.6} _{-0.7}$. The rest-frame yearly SN~Ia rate ($\textrm{SNR}_\textrm{Ia}$) is given by a two-component model 
 \citep{Scannapieco2005, Mannucci2005, Sullivan2006}         \begin{equation} 
   \label{abmode} 
       \left( \frac{\textrm{SNR}_\textrm{Ia}} { \textrm{yr}^{-1} }\right)  = \hat A \cdot 10^{-10}  \left( \frac{M_{\star}}{  \textrm{M}_{\odot} } \right)^{\alpha}   +\hat  B  \cdot 10^{-3}  \left(  \frac{ \textrm{SFR}}{  \textrm{M}_{\odot}   \textrm{yr}^{-1} }\right).
\end{equation}
We define the parameters $ \hat A =  1.05 \pm 0.16  $ and $\alpha = 0.68 \pm 0.01$ following \cite{smith_2011}. The parameter $ \hat B$ can be related to the
 $\textrm{SNR}_\textrm{cc} - \textrm{SFR}$  relation  
  \begin{equation} 
   \label{beta_k} 
    \hat  B = k_\textrm{cc} \Theta,
 \end{equation} 
where     $\Theta = \textrm{SNR}_\textrm{Ia} / \textrm{SNR}_\textrm{cc}$. At redshift $z < 1$, the ratio of $\textrm{SNR}_\textrm{Ia} / \textrm{SNR}_\textrm{cc}$   approximately ranges between $\Theta = 1/2$  and $\Theta = 1/4$, so we use  $\Theta = 1/4$ for the lower limit on our rates and $\Theta = 1/2$ for the upper limit.  The parameter  $k_\textrm{cc}$ can be determined by measuring the ratio between  $\textrm{SNR}_\textrm{cc} $ and $\textrm{SFR}$. Using observational data this ratio has been constrained to be
 $k_\textrm{cc} =  7.5 \pm 2.5  $ \citep{Scannapieco2005}. We use  $k_\textrm{cc} = 10  $ as the upper limit, and $k_\textrm{cc} =  5  $ as the lower limit. 

\begin{figure}
    \centering
    \includegraphics[width=\columnwidth]{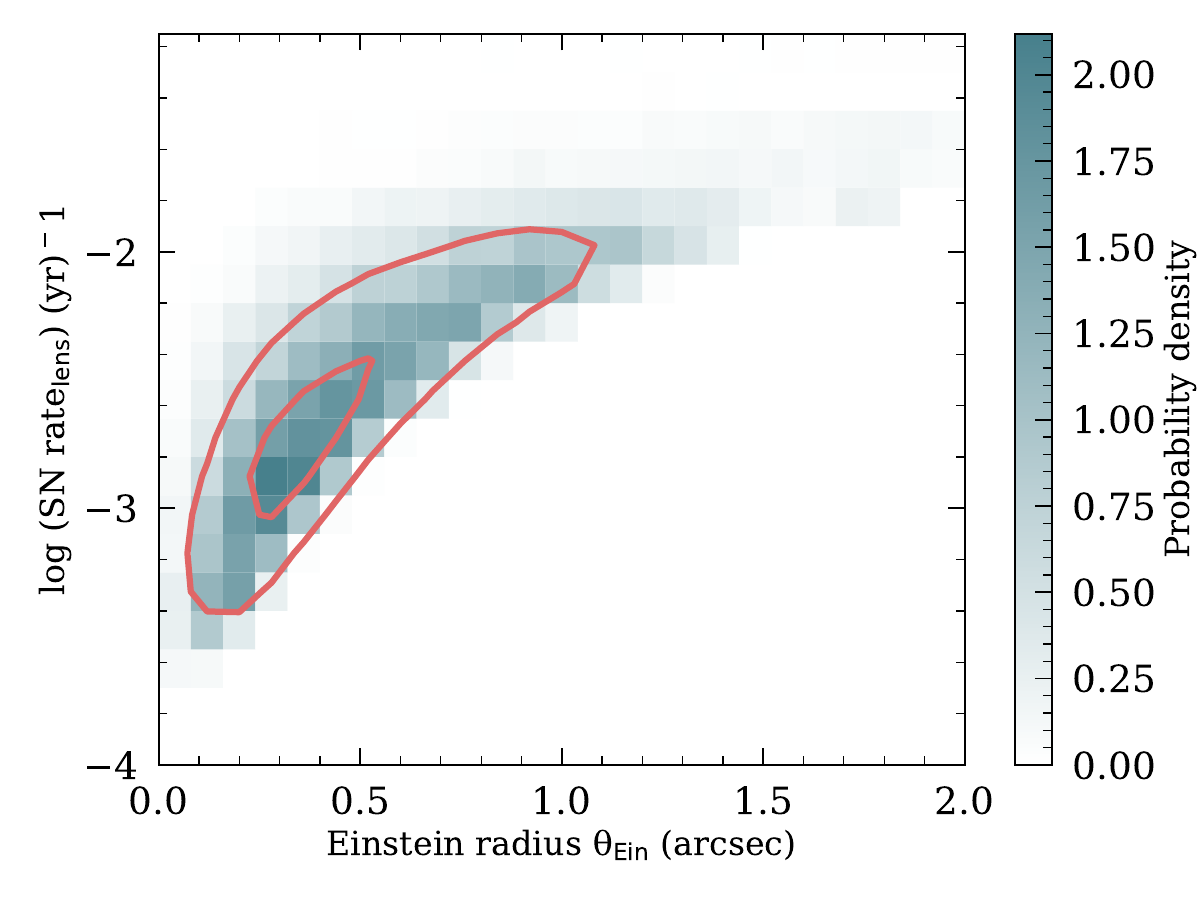}
    \caption{Rate of unlensed SN~Ia occurring in the lensing galaxy as a function of Einstein radius of the system.}
    \label{fig:SNe_in_lenses}
\end{figure}

As our lenses are red elliptical galaxies, we only focus on the rate component that is dependent of the stellar mass of the system. We find our lenses have a median SN~Ia rate of $\approx$ 0.005 SN~Ia yr$^{-1}$. For each lens, we use the observer-frame SN rate to estimate what fraction of SNe could actually be discovered by simulating a sample of $10^6$ SNe~Ia in the redshift range $z\in [0,1]$ and determining the fraction of these that would be discoverable. With our total detectable rate, given 10 000 lenses in LSST, the supernova rate in these lenses is of 229 SNe yr$^{-1}$. The rate of discoverable SNe  as a function of the lens Einstein radius is shown in Fig. \ref{fig:SNe_in_lenses}.

Following a similar method, we use Eqn. \ref{abmode} to calculate the SFR for every host galaxy. The host galaxies in our sample are less massive than our lenses, with median stellar mass log $M_{\star}/M_{\odot}$ = $9.9 ^ {+ 0.7} _{-0.9}$. These younger populations do undergo star-formation, with median log SFR = $0.0 ^{+ 0.6}_{-0.73}$. We find the median SN~Ia rate per host galaxy to be between 0.002-0.005 SN~Ia per rest-frame year, from our lower and upper limits respectively. We convert these to observer-frame rates and use our results from \texttt{simsurvey} to estimate the rate of glSNe in every host galaxy that will be discoverable.

We define the false positive rate (FPR) of this search method as 
\begin{equation}
    \mathrm{FPR} = \frac{1}{N_{\mathrm{sim}}}\sum_{i}^{\mathrm{N_{sim}}}\frac{\mathrm{FP}_{i}}{\mathrm{FP}_i + \mathrm{TP}_{i}},
\end{equation}
where $N_{\mathrm{sim}}$ is the number of lenses in our simulation. For each lensed system $i$, $\mathrm{FP}$ represents the observer-frame rate of detectable unlensed SNe~Ia in the lens and $\mathrm{TP}$ is the observer-frame rate of detectable glSNe~Ia in the source. We find FPR = 0.64 -- 0.77 depending on our rate parameters, which is consistent with the overall low probability of having a lensing event. Most SNe detections around lenses will actually correspond to unlensed transients in the lens galaxy. Relaxing our detection requirements to one 3$\sigma$ detection increases this FPR further to 0.84. Whilst such a false positive rate is likely tolerable given the high scientific value of lensed SNe, further cuts could be introduced to reduce the false positive rates, for example the position of the transient within the lensed system (first images form close to or beyond the Einstein radius, where the lens stellar density is lower) or the colour and magnitude at discovery.

\section{Limitations}
\label{sec:limitations}
The forecasts presented in this work are based on Monte Carlo simulations which, despite being well motivated by observations, are based on a series of assumptions that could bias our predictions. These should therefore be seen as a general estimate. In this subsection we comment on the main limitations of our work and what effect this could have on our predictions. 

\subsection{Source population}

Our simulations make use of the SCOTCH catalogue to establish correlations between properties of SNe and their host galaxies. \cite{SCOTCH} list a series of assumptions that were required to simulate the source catalogue. While they discuss in detail the implications of these assumptions, the most relevant one to this work is that the catalogue does not account for any kind of redshift evolution of the host galaxy properties. This is due to the fact SCOTCH itself is based on the GHOST catalogues \citep{GHOST}, which contain observational properties from ground-based surveys and are mostly complete up to redshift $z$ < 0.6 (but not strictly volume or redshift-limited to $z$ < 0.6). Thus, these catalogues probably do not accurately represent the supernova-host relationships beyond redshift $z$ > 0.6. Recent work by \cite{Popovic2024} shows evidence for evolving galaxy properties with redshift, which had already been hinted at in the past \citep{Nicolas2020} but was hard to determine partly due to Malmquist bias \citep{Popovic2021}. The discovery of larger samples of SNe at higher redshifts with upcoming deeper photometric surveys will allow us to confirm whether there exists some redshift evolution in host-transient correlations.

\subsubsection{Rates and luminosity functions}
Aside from the relationship between the host galaxy and the supernova, the properties of the supernova itself will also have an impact on our results. This becomes evident when considering our rate forecasts, which are roughly 
consistent with previous rate estimates (e.g. \citealt{Goldstein2019}, \citealt{Wojtak-19}, \citealt{Nikki_rates}, \citealt{SainzdeMurieta}). One exception is our rate estimates for SNe~IIn. Previous rate estimates find that this subtype of core-collapse supernovae will dominate glSN rates, due to their bright luminosity functions that extend to peak absolute magnitudes brighter than $M_{B} \gtrsim -19$. The rates of discoverable SNe rely heavily on a wide range of assumptions about the source population, including the SED and probability distribution of the intrinsic luminosity. 

\begin{figure}
    \centering
    \includegraphics[width=\columnwidth]{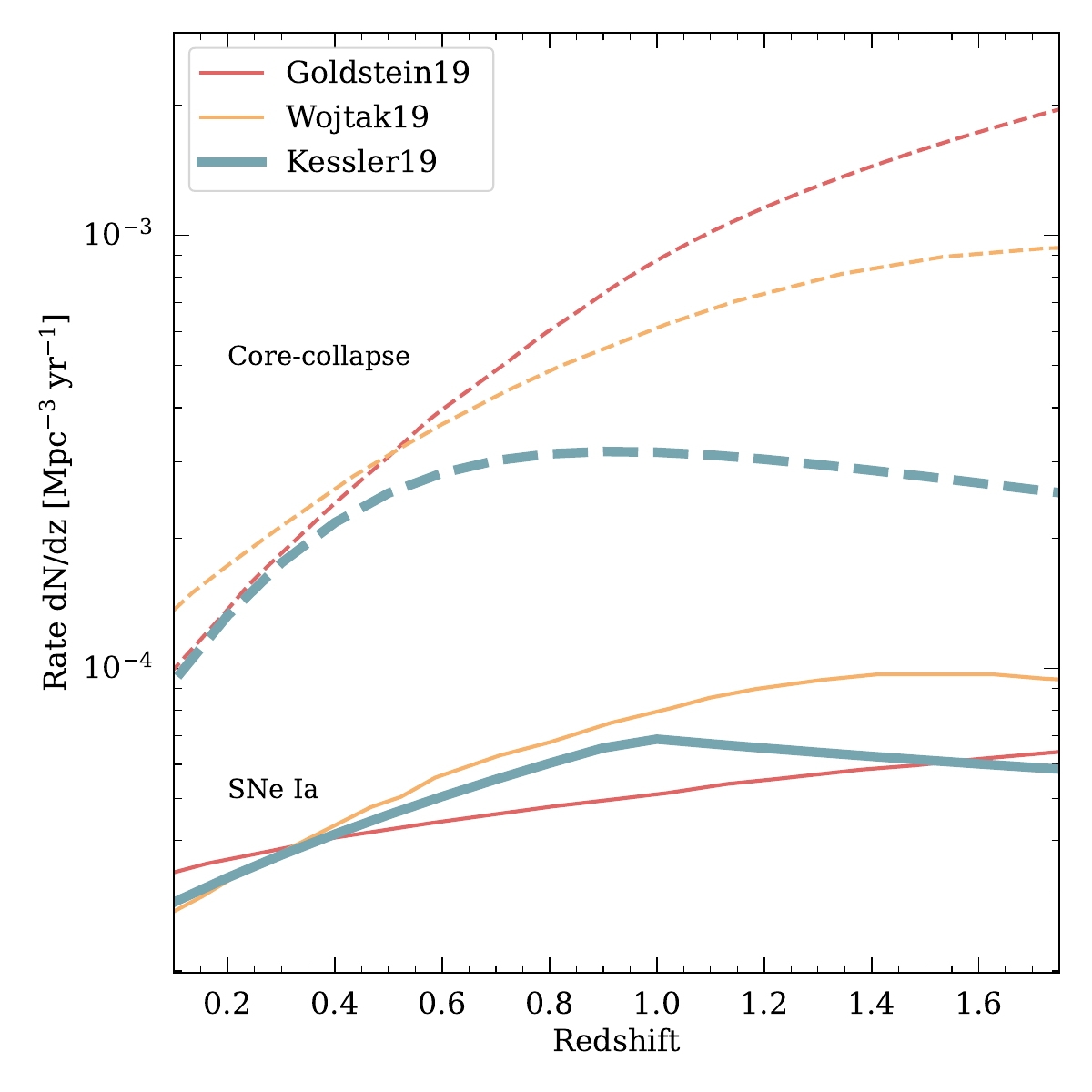}
    \caption{Volumetric SN rates used in this work (from \citealt{kessler2019}), compared with those used in \citealt{Wojtak-19} and \citealt{Goldstein2019}. The rates in this worked are represented by the blue lines. The rates we have used show a smaller number density of supernovae at high redshift.}
    \label{fig:SN_RATES}
\end{figure}

One of the key differences between our study and previous works is the choice of volumetric rates in our simulations. A comparison between the different rates used in different glSN yield studies are shown in Fig. \ref{fig:SN_RATES}. Core-collapse supernovae rates follow the star formation history (SFH) (e.g. \citealt{Madau_Dickinson_2014}). Throughout this work we used the same volumetric rates as \cite{kessler2019}, which are given by \cite{Strolger}. These results are consistent with the SFR densities derived from dust-corrected UV emission, but disagree with SN rates that follow the SFH evolution from \cite{Madau_Dickinson_2014}. \cite{Wojtak-19} use a model for core-collapse rates that is directly proportional to the star formation history by \cite{Madau_Dickinson_2014}. \cite{Goldstein2019} use the core-collapse volumetric rates by \cite{Li_2011}, which state an evolution of the rates $\propto (1+z)^{3.6}$. These rates show a very different evolution with redshift, and so a much larger number of core-collapse supernovae beyond redshift $z \sim 0.8$.

There also exist some discrepancies in the choice of volumetric SN~Ia rates. While we again use the volumetric rates from \cite{kessler2019}, based on two different observational studies of supernovae, \cite{Wojtak-19} used a more theoretical approach based on the star formation rate and the delay-time distribution. \cite{Goldstein2019} used the volumetric rates from \cite{Sullivan2006}. These rates, once again, show a higher number of SNe~Ia in the Universe. 
In practice, it is highly likely that none of our volumetric rate estimates are a completely accurate representation of the volumetric rate of supernovae in our Universe. The discovery of more glSNe at higher redshifts will allow for a better understanding of the redshift evolution of the volumetric rate. 

\begin{figure*}
    \centering
    \includegraphics[width=\textwidth]{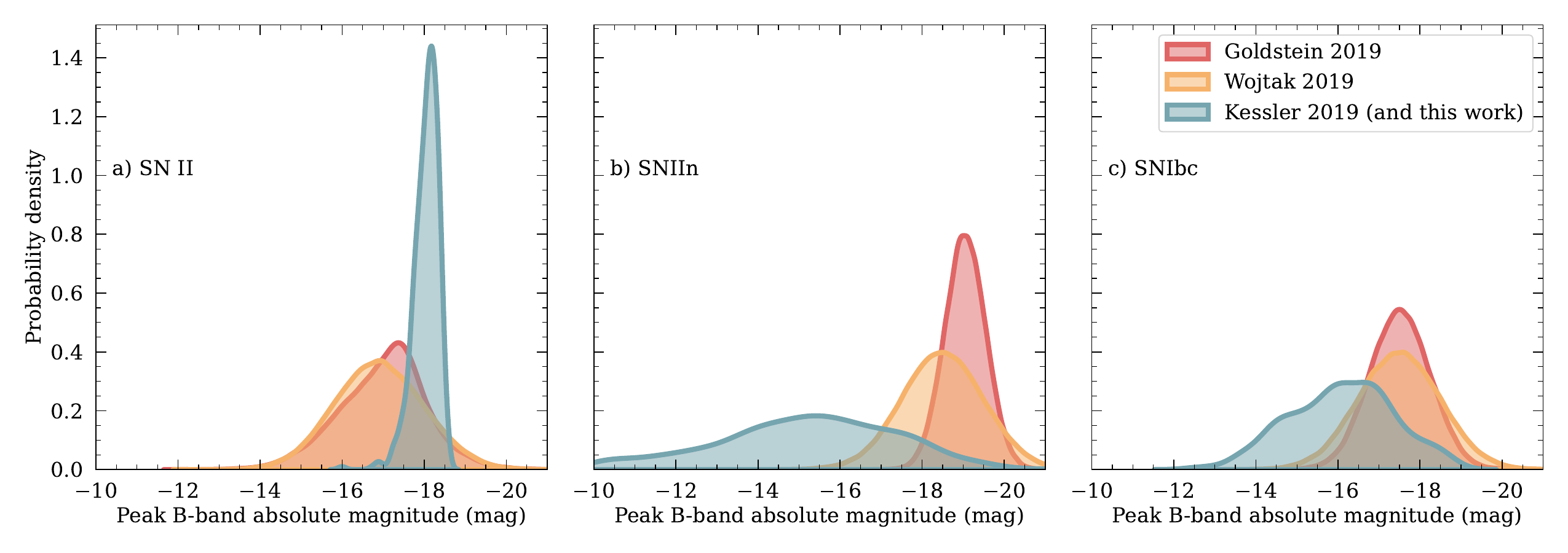}
    \caption{Luminosity function used in this work for the peak magnitude of different core-collapse SNe (based on \citealt{kessler2019}) compared with those used by \citealt{Goldstein2019} and \citealt{Wojtak-19}.}
    \label{fig:luminosity_fns}
\end{figure*}

Another difference is the choice of luminosity function, particularly for core-collapse SNe. Figure \ref{fig:luminosity_fns} shows the luminosity function for the SN templates in this work (based on \citealt{kessler2019}), \cite{Wojtak-19} and \cite{Goldstein2019}. The luminosity distribution of SN~II used in this work has a narrower spread, which means that even if the median peak magnitude of our population of SN~II is brighter, the ones in other studies extend to brighter absolute magnitudes. This is due to the templates used by \cite{kessler2019}. Conversely, our populations of SNIbc and SNIIn are significantly fainter, with differences reaching up to four magnitudes. These faint glSNe would require extremely high magnifications in order to push them beyond the detection threshold, which are more unlikely as they can only take place if the SN is located in a region near the caustic. It is therefore unsurprising that our rate estimates are much lower. 

In order to quantify the impact of luminosity functions and volumetric rate assumptions on rate forecasts, we change some of our initial assumptions and study their impact on rates. For SNe~Ia, which have better understood luminosity functions, we simply use different volumetric rates. Using the rates from \cite{Wojtak-19} results in a $\approx 15\% $ increase in glSNe~Ia rates. The use of volumetric rates from \cite{Goldstein2019} results in a 5\% decrease in yields. For SN~II, which dominate our rate forecasts, we need to take into account both the differences in rates and luminosity function. We re-run our simulations with the different assumptions chosen by \cite{Wojtak-19} and \cite{Goldstein2019}. As expected given Fig. \ref{fig:SN_RATES} and Fig. \ref{fig:luminosity_fns}, the use of these rates results in a large increase in rates. Using the assumptions from \cite{Goldstein2019} results in a 65\% increase in glSN~II rates. The assumptions in \cite{Wojtak-19} result in a 20\% increase in rates. Despite the differences in rate forecasts, their impact on our estimates for the fraction of lensed hosts is negligible.

These differences highlight the intrinsic uncertainty in glSN rate forecasts, and the importance of detailing the assumptions such forecasts are built upon. Upcoming deeper surveys will allow us to increase our sample of core-collapse supernovae, which will improve our understanding of their properties at higher redshifts.

\subsubsection{Scatter due to dust and microlensing}
Our simulations accounted for extinction from the Milky Way and dust in both the host and lens galaxies. The uncertainty coming from the distribution of dust in galaxies is a source of significant scatter, especially for core-collapse supernovae, which happen in star-forming and dusty environments \citep{Hatano_1998}. Along with their dimmer lightcurves, dust is believed to have a big impact on the lower rates of core-collapse supernovae \citep{Mannucci2003}. Our treatment of dust was the same for both SNe~Ia and core-collapse supernovae, therefore it is likely that we underestimated the effects of dust, making the core-collapse rates an upper limit. 

For simplicity, we did not account for the effects of microlensing. In most scenarios, this will slightly demagnify the glSNe, with some glSNe being more magnified (example histograms for the distribution of microlensing magnifications can be found in \citealt{Weisenbach2021, Weisenbach2024}). Some recent rate forecasts \citep{Nikki_rates, SainzdeMurieta} study the effect of microlensing on glSNe yields and find that for surveys with the depth of LSST, microlensing demagnification pushes some supernovae below the detection threshold, leading to an 8\% decrease in rates. We introduce a simple microlensing model in order to study whether this would have a significant impact on our rates. We re-simulated our glSN lightcurves adding a 20\% scatter in the flux of the unlensed model. This means some supernovae will be up to 20\% dimmer than originally simulated but others will be up to 20\% brighter. We find a 9\% decrease in our glSN estimates, which is consistent with \cite{Nikki_rates, SainzdeMurieta}. We also checked the effect of microlensing on our lensed host fraction estimates and found that accounting for microlensing had a negligible impact on our results.

\subsection{Deflector population}
Another source of uncertainty will come from the choice of lens model, which will have an effect on the time-delays and magnifications of our supernovae. Our lens population was simulated according to \citet{Collett2015}. We assumed SIE profiles for all lens galaxies, which has been shown to be a good approximation for massive elliptical galaxies \citep{Auger2009}. We did not include any kind of evolution on the velocity dispersion with redshift. This is consistent with results from \citet{Bezanson2011} and \cite{Shu2012}, at redshifts up to $z\leq 0.5$, where most of our lenses are found.  Recent work by \citet{Ferrami2024} has shown the redshift dependence of the velocity dispersion function of the deflector population can have an impact on strong lensing estimates. If we instead use the velocity dispersion function from \cite{Geng2021}, which displays a negative evolution of the number of deflectors with redshift, our forecast rate of glSNe~Ia decreases by $33\%$. However, because this velocity dispersion function removes high redshift lenses, we actually see a slight increase in the fraction of glSNe~Ia systems with a multiply imaged host: the fraction discoverable with LSST and/or Euclid increases by $8\%$. We conclude that the uncertainty on the $z>0.5$ velocity dispersion function produces substantial uncertainty on the forecast rate of glSNe, but it is less important for the rate of glSNe in discoverable lenses.

\subsection{Discoverability criteria}

Our actual ability to discover lenses, as well as our ability to identify glSNe will also have an impact on our rate estimates. In Section \ref{sec:lensproperties} we have already quantified the impact of our Einstein radius and magnitude cuts on the rate of lensed glSN hosts. We showed a difference of $\approx$ 20\% in the number of lensed hosts that are discoverable between our more optimistic and more pessimistic scenarios. 

Finally, we made the assumption that all the glSNe can be resolved from both the lens and the host, without loss of discovery depth. However, as shown in Fig. \ref{fig:discoverable_properties2}, our findings indicate that discoverable glSNe in galaxy-galaxy lenses generally occur closer to the centres of their hosts. As a result, some of these glSNe may not be detected in the LSST difference image analysis pipelines. Quantifying this effect is beyond the scope of this paper, but it is likely that this will decrease the rate of glSNe discovered in LSST.

\section{Conclusions}
\label{sec:conclusions}

In this work we have investigated the hosts of lensed SNe. We built a population of galaxy-scale lenses and put them in front of simulated SNe and their host galaxies using the SCOTCH catalogue \citep{SCOTCH}. We did this to answer two questions: how often are the hosts of glSNe also strongly lensed, and how many glSNe should we expect to discover by monitoring the population of galaxy-galaxy lenses that Euclid and LSST will soon deliver. We simulated SNe type Ia, Ibc, II, IIn and SLSNe. Our model predicted annual rates of 88 with LSST, with type II supernovae being the dominant subtype followed closely by type Ia supernovae.

A fraction of these glSNe, however, will not have a strongly lensed host. Thus, even if their time-delays are long enough to be measured precisely, it will not be possible to use them to infer the value of $H_0$, as lens modelling is needed to convert time delays into a value of the Hubble constant. We determined that in $54\%$ of cases, both glSNe and their host galaxy are lensed. $25\%$ of all glSNe should have a multiply imaged host that is sufficiently bright, magnified and resolved that they would be 
detectable as a galaxy-galaxy lens in at least one of the wide-field surveys of Euclid or LSST. Because such systems have larger Einstein radii, they typically have longer time delays and are more suited for cosmography.

Monitoring known galaxy-galaxy lenses can potentially allow us to be less strict in discovery thresholds since any change in these systems is either a lensed transient or a false positive transient in the lensing galaxy. Changing from three 5$\sigma$ detections to a single 3$\sigma$ detection would increase glSNe rates by 30\% as well as allow us to discover them 4 days earlier on average. Discovering glSNe early gives us more time to take follow-up with other instruments and study the early evolution of glSNe. 
One drawback of this approach is the false positive rate: 84\% of $3\sigma$ changes will be due to unlensed SNe in the lens galaxy.

Our simulations rely on assumptions regarding the mass profile of lenses, the volumetric rates of unlensed transients, and supernova lightcurve templates. We have quantified the effect of some of these assumptions and concluded that they have a bigger impact on rate estimates than on the fractions of glSNe with lensed hosts. The different choices likely explain why published forecast glSNe rates can differ by a factor of two \citep{SainzdeMurieta,Goldstein2019,Nikki_rates,Wojtak-19}.

Monitoring known gravitational lenses for new transients is an established field, but is currently limited by the number of known lenses. In the LSST era, only a quarter of all glSNe will occur in galaxy-galaxy lenses that could plausibly be discovered beforehand in Euclid and/or LSST, but this rises to a third for the glSNe that are golden for cosmology. Finding and monitoring large samples of galaxy-galaxy lenses can therefore give us a realistic pathway to $H_0$ with glSNe even if searching for glSNe in the full LSST alert stream proves challenging. We have found that 7 glSNe~Ia suitable for cosmology should be discoverable with LSST per year by monitoring the galaxy-galaxy lenses we expect to discover in LSST and Euclid. Achieving this yield should result in a 1.3\% precision measurement of $H_0$ after 10 years, assuming we have the galaxy-galaxy watch list in place at the start of the LSST survey.

\section*{Acknowledgements}
We are grateful to Nikki Arendse, Suhail Dhawan and Anupreeta More for helpful conversations that have enriched this work. This project has received funding from the European Research Council (ERC)
under the European Union’s Horizon 2020 research and innovation
programme (LensEra: grant agreement No 945536).

T.~E.~C. is funded by a Royal Society University Research Fellowship. 
This work is supported by the National Science Foundation under Cooperative Agreement PHY-2019786 (The NSF AI Institute for Artificial Intelligence and Fundamental Interactions, \url{http://iaifi.org/}).
\section*{Data Availability}
All data are simulated: our simulations are available from the corresponding author upon request.



\bibliographystyle{mnras}
\bibliography{glSNe_lenses}


\appendix

%

\bsp	
\label{lastpage}
\end{document}